\documentclass[aps,prl,reprint,superscriptaddress]{revtex4-2}
\usepackage{amsmath}
\usepackage{graphicx}
\usepackage{units}  
\usepackage{xspace}
\usepackage{subfig}
\usepackage{hyperref}

\newcommand{\ve}{\varepsilon}
\newcommand{\mb}{\mathbf}

\newcommand{\beq}{\begin{equation}}
\newcommand{\eeq}{\end{equation}}
\newcommand{\bea}{\begin{eqnarray}}
\newcommand{\eea}{\end{eqnarray}}
\newcommand{\mk}{\mathbf{k}}
\newcommand{\xik}{\xi_{\mathbf{k}}}
\newcommand{\tk}{t_{\mathbf{k}}}

\newcommand{\sn}{\sigma_\textrm{norm}}
\newcommand{\sa}{\sigma_\textrm{anom}}
\newcommand{\mr}{\mathrm}

\usepackage[usenames]{color}

\begin{document}

\bibliographystyle{apsrev4-2}
 
\title{Anomalous Surface Conductivity of Weyl Semimetals}

\author{Hridis K. Pal}
\affiliation{Department of Physics, Indian Institute of Technology Bombay, Powai, Mumbai 400076, India}
\author{Osakpolor Eki Obakpolor}
\affiliation{Department of Physics, University of Houston, Houston, TX 77204, USA}   
\author{Pavan Hosur}
\affiliation{Department of Physics, University of Houston, Houston, TX 77204, USA}   

\begin{abstract}
We calculate the surface dc conductivity of Weyl semimetals and show that it contains an anomalous contribution in addition to a Drude contribution from the Fermi arc. The anomalous part is independent of the surface scattering time, and appears at nonzero temperature and doping (away from the Weyl nodes), increasing quadratically with both with a universal ratio of coefficients. Microscopically, it results from the contribution of the gapless bulk to the surface conductivity. We argue that this can be interpreted as the conductivity of an effective interacting surface fluid that coexists with the Fermi arc metal. In a certain regime of low temperatures, the temperature dependence of the surface conductivity is dominated by the anomalous response, which can be probed experimentally to unravel the unusual behavior. 
\end{abstract}

\maketitle 

%Intro%

\emph{Introduction.} Weyl semimetals (WSMs) are gapless topological materials defined by non-degenerate bands in the bulk intersecting at isolated points in the momentum space \citep{VafekDiracReview,Burkov2018,Burkov:2016aa,YanFelserReview,ArmitageWeylDiracReview,Shen2017,Belopolski:2016wu,Guo2018,Chang2016,Gyenis_2016,Huang:2015vn,Inoue1184,Lv:2015aa,Sun2015a,Xu2015,Xu2016,Xu613,Yang:2015aa,Zheng2016}. These points, known as Weyl nodes, have a well-defined chirality and host quasiparticles that mimic relativistic Weyl fermions first studied in high-energy physics. Since Weyl fermions are topological objects, their response to external electromagnetic fields is distinct from that of usual Bloch electrons in conventional solids. Numerous such  bulk topological responses have been extensively investigated, both in theory and in experiments, over the past decade \citep{Hosur2013a,Wang_2018,Hu:2019aa,ZyuninBurkovWeylTheta,ChenAxionResponse,VazifehEMResponse,Burkov_2015,Hosur2012,Juan:2017aa,Wang2017,Halterman2018,Halterman:19,Nagaosa:2020aa,NielsenABJ,IsachenkovCME,SadofyevChiralHydroNotes,Loganayagam2012,GoswamiFieldTheory,Wang2013,BasarTriangleAnomaly,LandsteinerAnomaly}.

The surface of a WSM, however, has remained more mysterious. It hosts unusual metallic states known as Fermi arcs (FAs) that connect surface projections of bulk Weyl nodes of opposite chiralities. They do not form a closed contour, unlike Fermi surfaces in conventional two-dimensional (2D) metals \citep{Benito-Matias2019,Chang2016,Gyenis_2016,Deng2017,Deng:2016aa,Guo2018,HaldaneFermiArc,Hosur2012a,Huang2016,Huang:2015vn,Iaia:2018aa,Inoue1184,Kwon:2020aa,Lau2017,Sakano2017,Sun2015a,Lv:2015aa,Xu2015,Xu2015a,Xu2015b,Xu2016,Xu613,XuLiu2018,Yuan2018QPI,Yuaneaaw9485,Zhang:2017ac,Moll:2016aa,Potter2014,Zhang2016}. This raises the possibility of unusual response on the surface to external electromagnetic fields \citep{Gorbar2016,Wilson2018,Resta2018,Mukherjee2019,Breitkreiz2019}. However, investigations in this direction are rendered challenging by the fact that at the end points of the FAs, the wavefunction merges with the bulk Bloch waves at the Weyl nodes and renders the surface of a WSM fundamentally inseparable from the bulk. This is in stark contrast to the surface states of topological insulators, which decay exponentially into the bulk at all surface momenta \cite{HasanKaneReview, QiZhangRMP}. The bulk-surface inseparability in WSMs can also be understood from an energy-based perspective. Since a topological insulator is gapped in the bulk, its low-energy theory consists of a strictly surface Hamiltonian. In contrast, WSMs have gapless states both in the bulk and on the surface, so an energy cut off is not available to disentangle the surface from the bulk. While the surface-bulk inseparability promises rich physics such as quantum oscillations from cyclotron orbits composed of FAs and chiral modes in the bulk \citep{Borchmann2017,Moll:2016aa,Potter2014,Zhang2016,Wang2017a,Zhang:2017ac,Zhang2019trefoil,Nishihaya:2021um,Zhang:2019ab}, unusual collective modes \citep{Bonacic2018,Bugaiko2020,Gorbar2019,Hofmann2016,Tamaya_2019,Song2017,Andolina2018,Adinehvand2019,Ghosh2020} and dissipative chiral transport \cite{Gorbar2016}, it invalidates a strictly surface Hamiltonian, thereby hindering a controlled theoretical description of the surface and leaving its electromagnetic response poorly understood.

In this work, we calculate a basic surface transport property of a WSM, namely, the dc conductivity $\sigma_\textrm{surf}$. Unlike previous studies of surface transport in WSMs that exclusively focus on the contributions of the FAs \citep{Gorbar2016,Wilson2018,Resta2018,Mukherjee2019,Breitkreiz2019}, we include the nontrivial influence of the bulk on the surface in a controlled manner and show that $\sigma_{\textrm{surf}}$ is comprised of two qualitatively different contributions:
\beq
\sigma_\textrm{surf}=\sn + \sa.
\label{sigtotal}
\eeq
Here, $\sn$ is a normal, Drude-like conductivity arising from the FA. It is proportional to the scattering time ($\tau$) and has a negligible dependence on temperature ($T$) besides the $T$-dependence of $\tau$. In contrast, $\sa$ is an anomalous term arising from the bulk states. It is $\tau$-independent, but has a characteristic dependence on both $T$ and doping around the Weyl nodes, $\mu$: $\sa\propto\mu^2+k_B^2T^2\pi^2/3$. Crucially, the coefficients have a universal ratio, $k_B^2\pi^2/3$, independent of material parameters.  
In a regime of low temperatures described later, the $T$-dependent part of $\sigma_\textrm{surf}$ is dominated by $\sa$ which provides immediate experimental access to $\sa$. 
Note that $\sa$ is $\tau$-independent even if the bulk quasiparticles have a nonzero lifetime. This is reminiscent of the universal minimal conductivity of a 2D Dirac fermion that only depends on fundamental constants \cite{KATSNELSON20073}. As espoused later, the electrical response of the bulk states on the surface mimics that of an interacting 2D fluid that coexists with the FA metal on the surface. This, in turn, gives rise to unusual form of $\sa$ in Eq.~(\ref{sigtotal}). Thus, the daunting problem of the 2D FAs coexisting with the 3D bulk states reduces to a tractable, strictly 2D, effective two-fluid problem.

%% Model %%

\emph{Model.} We employ a variant of the model introduced in  Ref.~\onlinecite{Hosur2012a} for generating a WSM with arbitrary configurations of FAs and Weyl nodes. The model consists of a finite stack of metallic bilayers whose Hamiltonian as a function of $\mathbf{k}=(k_x,k_y)$ is:
\beq
H_{\mk}=\sum_{z=1}^L\psi_{z,\mk}^{\dagger}[(-1)^z\xi_{\mk}-\mu]\psi_{z,\mk}-\sum_{z=1}^{L-1}\psi_{z,\mk}^{\dagger}t_{z,\mk}\psi_{z+1,\mk}+\mathrm{H.c.},
\label{ham}
\eeq
where $\psi_{z,\mk}^\dagger$ creates an electron with momentum $\hbar\mk$ in layer $z$, $\xi_{\mk}$ denotes a metallic dispersion with Fermi surface along $\xi_\mk = 0$, $\mu$ is the bulk chemical potential and the interlayer hopping modulates as $t_{z,\mk}=t_\perp +(-1)^z\delta t_\mk$. If $|\delta\tk|<|t_\perp|\ \forall\ \mk$, $H_\mk$ produces bulk Weyl nodes in the $k_z=\pi/2c$ plane, where $2c$ is the lattice constant along $z$, at discrete points $\mathbf K_j\equiv(K_{jx},K_{jy})$ such that $\xi_{\mathbf K_j}=\delta t_{\mathbf K_j}=0$. Near the $j^{th}$ node, the bulk Hamiltonian is $H_{\text{Weyl},j}(\mathbf{q})=\hbar\mathbf{q}\cdot\left(\mathbf{v}_{j}\sigma_{z}+\mathbf{u}_{j}\sigma_{x}+\mathbf{w}_{j}\sigma_{y}\right)-\mu$, where $\sigma_{\alpha}$ are bilayer Pauli matrices and 
$\mathbf{v}_{j}=\frac{1}{\hbar}\boldsymbol{\nabla}_{\mathbf{K}_{j}}\xi_{\mathbf{K}_{j}}$,
$\mathbf{u}_{j}=\frac{2}{\hbar}\boldsymbol{\nabla}_{\mathbf{K}_{j}}\delta t_{\mathbf K_j}$ 
and $\mathbf{w}_{j}= -\frac{2}{\hbar}t_\perp c\hat{\mathbf{z}}$ are Weyl velocities.

\begin{figure}
\centering
\includegraphics[width=0.9\columnwidth]{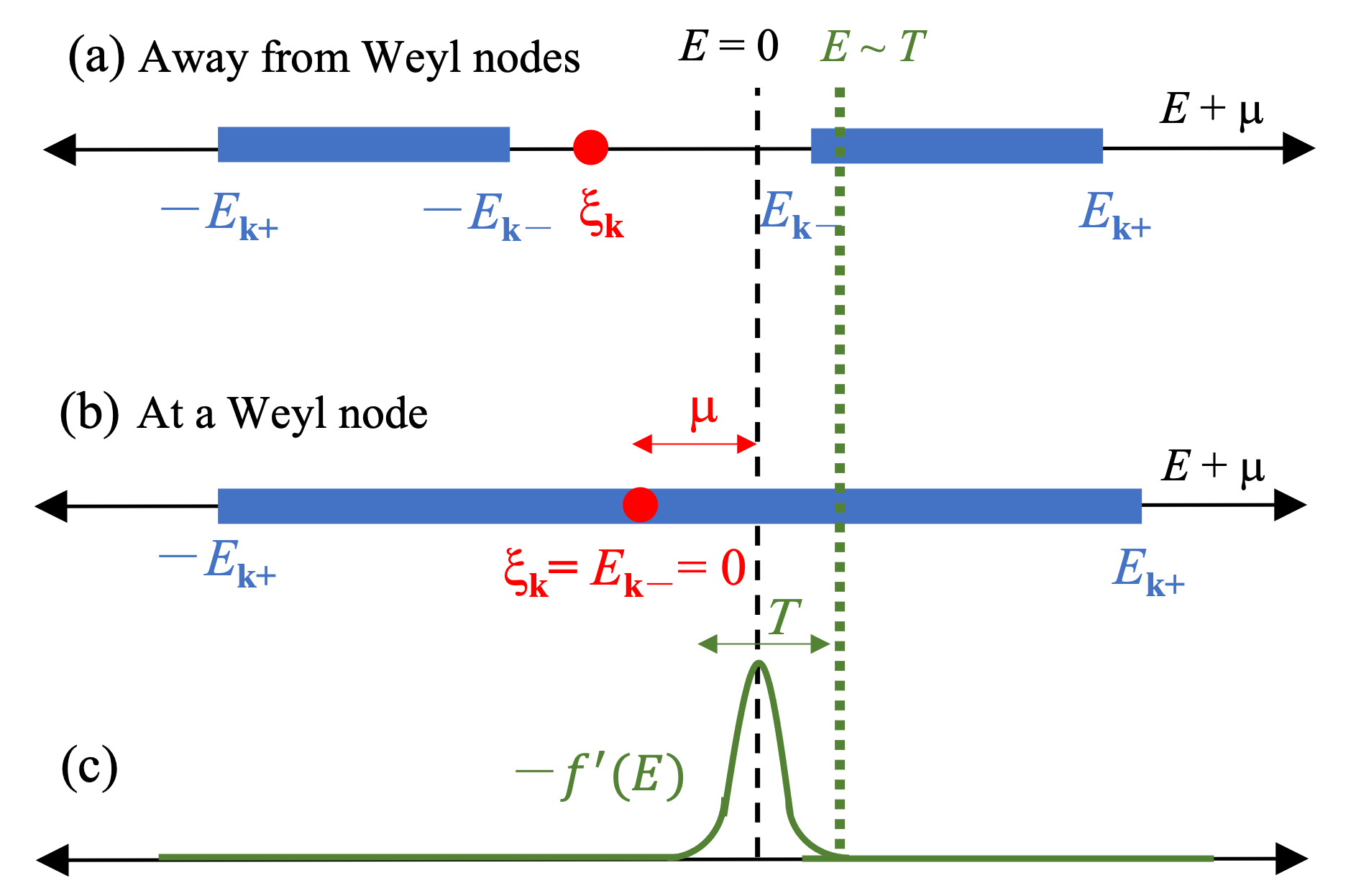}
\caption{Schematic of contributions to $\sa$. Red dots (blue bars) denote the pole (BCs) of $G_\mk(E)$ and physically represent the FA (bulk) states. (a) Away from the Weyl nodes, the BCs  do not contain $E=0$, resulting in no contribution to $\sa$ (b) At a Weyl node, $\xi_\mk=E_{\mk-}=0$, so both BCs touch the pole and $E=0$ lies within one of them, which results in a non-zero $\sa$ governed by $\mu$. (c) $T\neq0$ leads to $\sa$ by enabling access to states within a BC via broadening of $-\partial_Ef$.}
\label{schematic}
\end{figure}

The surface ($z=1$) Matsubara Green's function in the clean, non-interacting limit when $L\to\infty$ is \citep{obakpolor2021surface,Hosur2012a} (see Supplemental Material [SM] \cite{SM} for details)
\begin{eqnarray}
G_{\mk}(iE_n)&=&\frac{a_{\mk}(iE_n)+b_{\mk}(iE_n)}{2(t_\perp+\delta\tk)^2(iE_n+\mu-\xik)},\label{G11}\\
a_{\mk}(iE_n)&=&(iE_n+\mu)^2-\xik^2+4t_\perp\delta t_\mk,\nonumber\\
b_{\mk}(iE_n)&=&\sqrt{\prod_{\lambda=\pm}\left[(iE_n+\mu)^2-E_{\mk\lambda}^2\right]},\nonumber
\end{eqnarray}
where $E_{\mk+}=\sqrt{\xik^2+4t_\perp^2}$ and $E_{\mk-}=\sqrt{\xik^2+4\delta t_\mk^2}$ are the extremeties of the bulk conduction band at $\mk$. For real energies, $G_{\mk}(E)$ has BCs when $|E+\mu|\in[E_{\mk-},E_{\mk+}]$, i.e., $E$ lies within the bulk bands, which we dub the BC-region. When $|E+\mu|\not\in[E_{\mk-},E_{\mk+}]$, dubbed the N-region, $G_{\mk}(E)$ has simple poles along $E=\xik-\mu$ when $t_\perp\delta\tk>0$ that define the FAs (Fig.~\ref{schematic}). Importantly, the poles and BCs are generic features of surface Green's functions in WSMs and are not specific to the current model \cite{obakpolor2021surface}.

%General conductivity%

\emph{Surface conductivity.} Since our goal here is to investigate surface transport, we need to introduce a nonzero quasiparticle lifetime $\tau_\mk$ into Eq.~(\ref{G11}) that captures the effect of scattering required to relax momentum gained due to the external field. To this effect, we revisit Eq.~(\ref{ham}): a nonzero lifetime is introduced in each layer and the surface ($z=1$) Matsubara Green's function is reevaluated---see SM \cite{SM} for details. We assume scattering to be weak such that $1/\tau_\mk\rightarrow 0$. Analytically continuing, the retarded Green's functions in the N- and BC-regions are found to be
\begin{eqnarray}
G^R_{\mk}(E)&=&\frac{a_{\mk}(E)+b_{\mk}(E)}{2(t_\perp+\delta\tk)^2(E+\mu-\xik+ \frac{i\hbar}{2\tau_\mk})}, E\in N\label{gn}\\
G^R_{\mk}(E)&=&\frac{a_{\mk}(E)+ i\ \mathrm{sgn}(E+\mu)|b_{\mk}(E)|}{2(t_\perp+\delta\tk)^2(E+\mu-\xik+ \frac{i\hbar}{2\tau_\mk})}, E\in BC\label{gbc}
\end{eqnarray}
Note that we have assumed only $E$-independent scattering processes for simplicity, but relaxing this assumption will not change our qualitative results. 

The longitudinal dc conductivity along $x$-direction due to the motion of electrons only on the surface is given by
\beq
\sigma_\textrm{surf}=\frac{e^2\hbar}{2}\intop_{E,\mk}\left(-\frac{\partial f}{\partial E}\right)\left[v_{\mk,x}A_{\mk}(E)\right]^2,
\label{dccon}
\eeq
where $f(E)=1/[1+\exp(E/k_BT)]$, $\intop_{E,\mk}=\int\frac{d^2kdE}{(2\pi)^3}$, $v_{\mk,x}=\frac{1}{\hbar}\partial_{k_x}\xi_\mb{k}$ is the $x$-component of the in-plane velocity,  and $A_\mk(E)=-2\textrm{Im}G^R_\mk(E)$ is the surface spectral function. Due to the distinct forms of $G^R_\mk(E)$ in the N- and BC-regions, it is convenient to split the $E$-integral as $\int_{E}=\int_{E\in N}+\int_{E\in BC}\equiv\int_N+\int_{BC}$. The two terms yield $\sn$ and $\sa$ respectively, which leads to the decomposition of $\sigma_\textrm{surf}$ stated in Eq.~(\ref{sigtotal}). Moreover, in the limit $1/\tau_\mk\to0$, $\int_N$ receives contributions only from a sharp quasiparticle peak in $A_\mk(E)$ due to the FAs while $\int_{BC}$ depends only on a broad feature in $A_\mk(E)$ that captures the surface projection of the bulk states. Thus, $\sn$ and $\sa$ are in one-to-one correspondence with the FA and the bulk contributions to surface transport.

%sigma_norm%

In the N-region, Eqs.~(\ref{G11}), (\ref{gn}) and (\ref{dccon}) give $A_\mk(E)=2\pi W_\mk\delta_{\hbar/2\tau_\mk}(E+\mu-\xik)$, where   $W_\mk=R\left[\frac{4t_\perp\delta t_\mk}{(t_\perp+\delta t_\mk)^2}\right]$ with $R(x)=x\Theta(x)$ and $\delta_\eta(x)=\frac{1}{\pi}\frac{x}{x^2+\eta^2}$. For $1/\tau_\mk\to 0$, Eq.~(\ref{dccon}) then yields,
\beq
\sn=e^2\int_{N}\left(-\frac{\partial f}{\partial E}\right)\intop_\mk 2\pi\tau_\mk v_{\mk,x}^2W_\mk^2\delta(E+\mu-\xik).
\label{sigmaarcint}
\eeq
The factor $W_\mk^2$ above ensures that $\sn$ gets contributions only from the metallic FAs. Additionally, the contribution varies as one traverses the FA: it is largest near the middle of the FA (maximum $t_\perp\delta\tk$), vanishes at the Weyl nodes ($\delta\tk=0$), and remains zero in regions lacking FAs ($t_\perp\delta\tk<0$). At $T=0$, we find
\beq
\sn=2e^2 \nu_FD,
\label{sigmaarc}
\eeq
where $\nu_F=\intop_\mk {W}_\mk\delta(\mu-\xik)$ and $D=\frac{1}{2}\langle v_{F,x}^2\tau\rangle=(2\nu_F)^{-1}\intop_\mk\delta(\mu-\xik) (v_{\mk,x}{W}_\mk)^2\tau_\mk$ 
are the suitably weighted means of the density of states per unit area (DOS) and the diffusion constant due to the FAs. Clearly, $\sn$ is the Drude conductivity, scaling linearly with $\tau$. At nonzero $T$, temperature enters the conductivity mainly through $\tau_\mk$ which is model-dependent, and is discussed later.  Temperature may also enter via corrections to the DOS arising from the Sommerfeld expansion. However, such corrections vanish exactly for an ordinary 2D parabolic dispersion. By extension, the corrections are expected to be small for generic 2D metals including the FA metal, which we ignore henceforth.

\begin{figure}
\centering
\includegraphics[width=0.9\columnwidth]{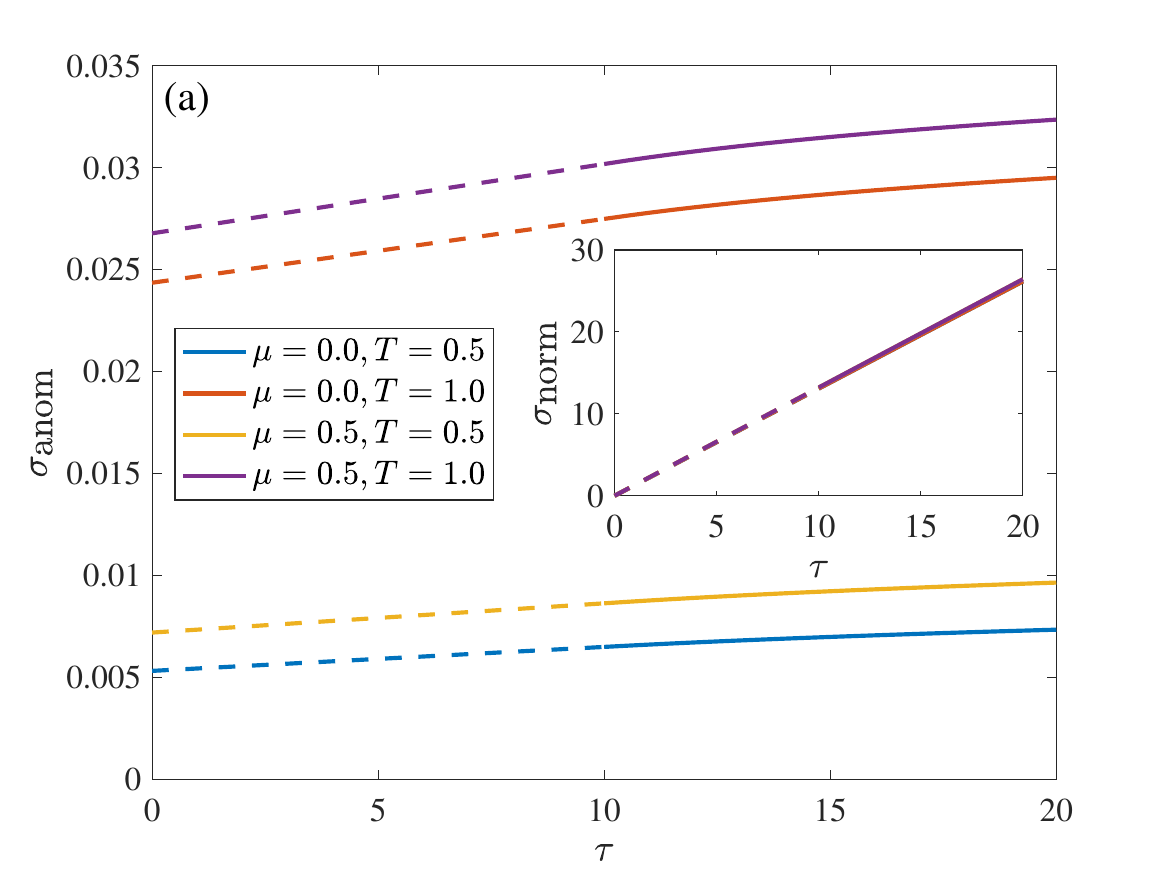}
\includegraphics[width=0.9\columnwidth]{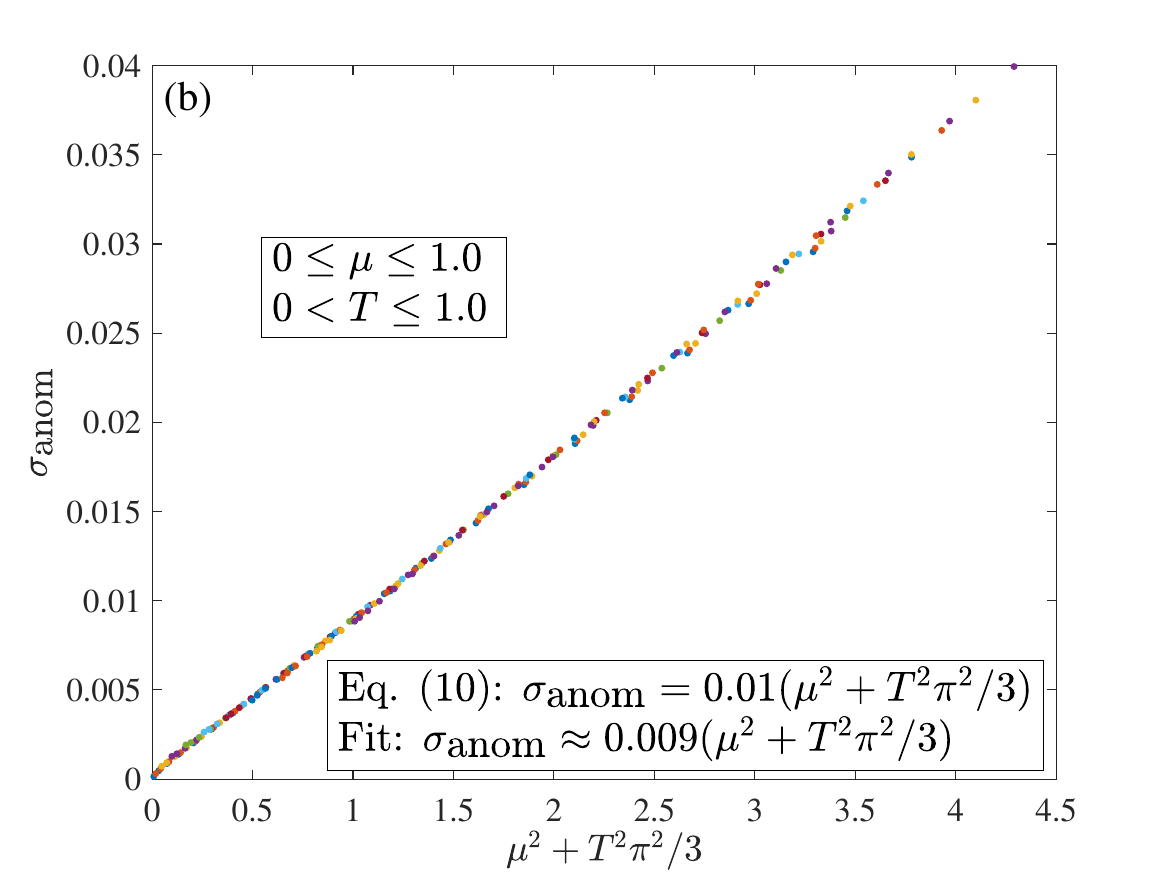}
\caption{$\sa$ and $\sn$ calculated numerically from Eq.~(\ref{dccon}) by separating the pole and BC contributions to $A_\mk(E)$ and assuming a $\mk$-independent lifetime $\tau$ for simplicity. (a) $\sigma_\textrm{anom}$ is nearly $\tau$-independent and has a finite intercept when extrapolated linearly to $\tau=0$, whereas $\sigma_\textrm{norm}\propto\tau$ (inset). (b) At the longest lifetime, $\tau=20$, a clear scaling collapse occurs: $\sigma_\textrm{anom}\propto\mu^2+T^2\pi^2/3$ with almost the precise pre-factor in accordance with Eq.~(\ref{sigmabcT}). We have used $\xik=k^2/2-50$, $t_\perp=10$ and $\delta\tk=-5k_y/k$, which gives a semicircular FA along $k=10$ for $k_y<0$, and set $|e|=\hbar=k_B=1$.}
\label{figsigdc}
\end{figure}

%Sigma_anom%

In the BC-region, Eqs.~(\ref{G11}), (\ref{gbc}) and (\ref{dccon}) give the anomalous contribution to the conductivity:
\begin{eqnarray}
\sa&=&2e^2\hbar\int_{BC}\left(-\frac{\partial f}{\partial E}\right)\intop_{\mk}v_{\mk,x}^2\times \nonumber\\
&& \frac{\left[(E+\mu)^2-E_{\mk-}^2\right]\left[E_{\mk+}^2-(E+\mu)^2\right]}{(E+\mu-\xik)^2(t_\perp+\delta\tk)^2}.
\label{sigbcT}
\end{eqnarray}
At low $T$, $f'(E)$ restricts $E$ to within $\pm\mathcal{O}(k_BT)$, so the BC-region effectively obeys $E_{\mk-}\lesssim k_BT+|\mu|\lesssim E_{\mk+}$. Now, $E_{\mk-}\geq0\ \forall\ \mk$ and vanishes only at the Weyl nodes, where the bulk bands touch. As a result, the dominant contribution to $\sa$ comes from the vicinity of the Weyl nodes. Thus, we can linearize around the $j^{th}$ node as $\xik\approx\hbar\mathbf{v}_j\cdot\mathbf{q}$ and $\delta\tk\approx \frac{\hbar}{2}\mathbf u_j\cdot\mathbf q$. Eq.~(\ref{sigbcT}) then evaluates to (see SM \cite{SM} for details)
\beq
\sa=\frac{e^2c}{\hbar^2\pi^2}\sum_j\left(\mu^2+\frac{k_B^2T^2\pi^2}{3}\right)\frac{v_{j,x}^2}{t_\perp|(\mathbf u_j\times\mathbf v_j)\cdot\mathbf w_j|},
\label{sigmabcT}
\eeq
where we assumed $|\mu|,k_BT\ll t_\perp$. If the Weyl nodes are at different energies, we must replace $\mu\to\mu_j$, the chemical potential relative to the energy of the $j^{th}$ node.  A remarkable feature of $\sa$ is that it is independent of $\tau$, the in-plane scattering time. Thus, $\sa$ manifestly has a different origin and expression compared to the Drude contribution, $\sn$. An exact numerical calculation of Eq.~(\ref{dccon}), shown in Fig.~\ref{figsigdc}, corroborates our analytical results for both the Drude as well as the anomalous conductivities. 
 
%Interacting surface fluid%

\emph{Effective interacting surface fluid.} We now reinterpret $\sa$ as the conductivity of an effective interacting surface fluid. To this end, we consider the surface particle density at $1/\tau_\mk\to0$, $n = \intop_{\mk,E}A_\mk(E)=n_\textrm{norm}+n_\textrm{anom}$ where $n_\textrm{norm} = \intop_\mk f(\xik-\mu)W_\mk$ and
\begin{equation}
n_\textrm{anom} = \intop_{\mk}\intop_{BC} f(E)\frac{\textrm{sgn}(E+\mu)|b_\mk|}{(t_\perp+\delta\tk)^2(E+\mu-\xi_\mk)}.
\end{equation}
As in the case of $\sa$, the integral receives contributions only from the regions near the Weyl nodes. Carrying out this integral by linearizing around the Weyl nodes as before, we get $n_\textrm{anom}=\sum_jn_{\textrm{anom},j}$, with
\begin{equation}
n_{\textrm{anom},j} =\frac{\mu\left(\mu^2+k_B^2T^2\pi^2\right)}{6\pi^2\hbar^3}\frac{c}{|(\mathbf u_j\times\mathbf v_j)\cdot\mathbf w_j|}.
\end{equation}
This, in turn, gives an effective DOS from the BC, $\nu_{\textrm{anom},j}=dn_{\textrm{anom},j}/d\mu$, in terms of which
Eq.~ (\ref{sigmabcT}) can be expressed as
\begin{equation}
\sa = \sum_j e^2\nu_{\textrm{anom},j}v_{j,x}^2\frac{2\hbar}{t_\perp}.
\label{sigmabc-dos}
\end{equation}
Written as above, $\sa$ resembles the Drude conductivity of a 2D metal, similar to $\sn$ in Eq.~(\ref{sigmaarc}), except that the DOS on the FA is replaced by an effective DOS on the surface due to the bulk and $\tilde\tau=2\hbar/t_\perp$ plays the role of lifetime in lieu of $\tau_\mk$. 

The appearance of a new effective lifetime can be traced to the fact that $\textrm{Im}G^R\neq0$ when $E\in BC$ even when $1/\tau_\mk=0$. In fact, one can write Eq.~(\ref{gbc}) as 
\begin{eqnarray}
    G^R_\mk(E)&=&\frac{Z_\mk(E)}{E+\mu-\xik+\frac{i\hbar}{2\tilde\tau_\mk(E)}},\nonumber\\
Z_\mk&=&\frac{a_\mk(E)^2+|b_\mk(E)|^2}{2a_\mk(E)(t_\perp+\delta\tk)^2} ,\nonumber \\
 \frac{\hbar}{2\tilde\tau_{\mk}(E)}&=&\mathrm{sgn}(E+\mu)(E+\mu-\xik)\frac{|b_\mk(E)|}{a_\mk(E)},
  \label{selfen}
\end{eqnarray}
in the region $|E+\mu|>E_{\mk -}$. This resembles an interacting Green's function where $Z_\mk(E)$ is the quasiparticle residue and $\tilde\tau_{\mk}(E)$ is the effective lifetime. At a Weyl node at $E=0$, we find $\tilde\tau_\mk(0) = \hbar/4t_\perp$ which shows up as the scattering time in the anomalous contribution upto pre-factors. Thus, the bulk states effectively induce an interacting 2D fluid on the surface near the Weyl nodes, leading to the additional anomalous contribution $\sa$ on the surface.

\begin{figure}
\centering
\includegraphics[scale=0.40]{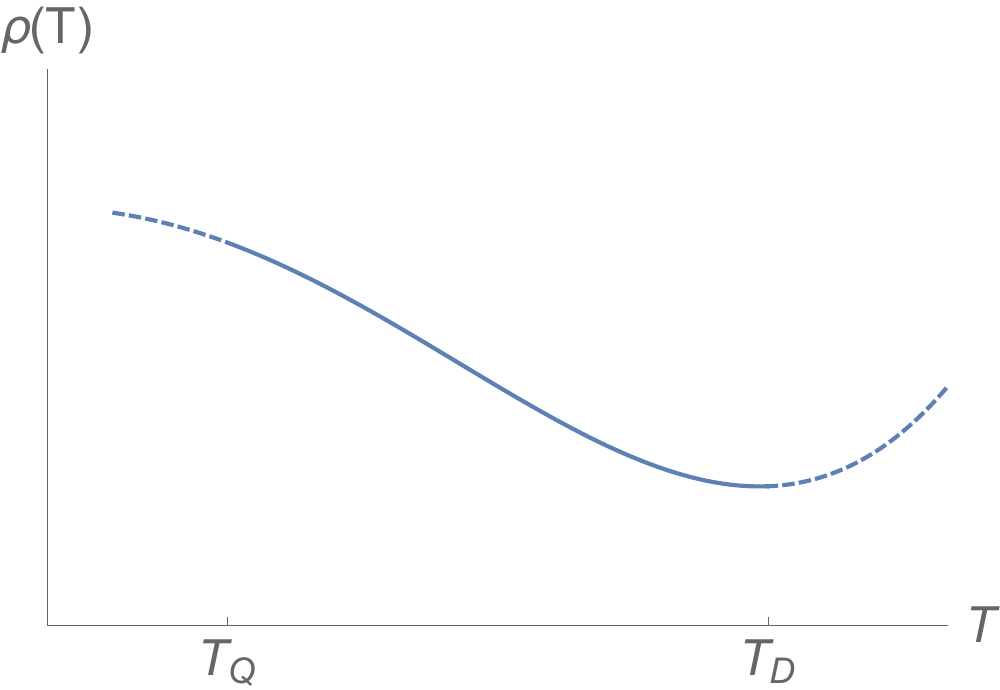}
\caption{Schematic depiction of the $T$-dependence of the surface resistivity, $1/\sigma_\text{surf}$, according to Eq.~(\ref{sigtotal}). Above $T_D$, the $T$-dependence of $\tau$ is significant whereas quantum corrections not considered in this work become important below $T_Q$. For $T_Q\lesssim T\lesssim T_D$, the resistivity is expected to grow quadratically with decreasing $T$ due to $\sa$.}
\label{resistivity}
\end{figure}

%Experiment

\emph{Experiment.} Our main results, Eqs.~(\ref{sigtotal}), (\ref{sigmaarc}), and (\ref{sigmabcT}), imply that the surface of a WSM, in spite of being a metal, exhibits a nonmetallic signature in transport: the conductivity (resistivity) increases (decreases) with increase in $T$ with a characteristic quadratic dependence. This can be studied experimentally to verify our theory. As an example we consider the commonly studied Weyl semimetal TaAs: for the W1 Weyl pockets, in-plane velocity $\approx 3\times 10^5\text{ m/s}$, out-of-plane velocity $\approx 3\times 10^4 \text{ m/s}$, $\tau\sim 0.1\text{ps}$, $c\sim 10\AA$, $t_{\perp}\sim 0.01\text{eV}$, and $\nu_F\sim 0.002-0.02\text{ eV}^{-1}\AA^{-2}$, depending on the size of the FA \cite{huang2015weyl,ramshaw2018quantum}. This yields $\sn\sim 10^{-4}-10^{-3} \Omega^{-1}$ and $\sa(T)-\sa(0)\sim 10^{-9}T^2\Omega^{-1}\text{K}^{-2}$, which evaluates to $\sim10^{-7}\Omega^{-1}$ for $T\sim10^1 \text{K}$. A change in resistance of $1$ part in $10^3-10^4$ is easily measurable, implying that our predicted effect is readily amenable to experimental verification; a similar estimate can be obtained for other Weyl materials. In arriving at this, however, we assumed 
$\tau$ to be $T$-independent in Eq.~(\ref{sigmaarc}). In reality, this is not true --- $\tau$ does depend on $T$ and is, in fact, the chief source of $T$-dependence in the $\sn$, which can compete with the $T^2$-dependence in $\sa$. Nevertheless, below a certain temperature $T_D$ whose exact value will depend on the microscopics, the $T$-dependence of $\sigma_\text{surf}$ can be dominated by $\sa$ even with a $T$-dependent $\tau$.  For example, if one assumes the usual form of electron-phonon scattering with the scattering time given by $\hbar/\tau_{e-ph}\sim k_BT^4/T_{BG}^3$, ($T_{BG}$ is the Bloch-Gruneisen temperature for the FA states), then $T_D\sim (\hbar k_B^{1/2}/\tau)(T_{BG}/t_{\perp})^{3/2}$. For $T_{BG}\sim100\text{ K}$ and $t_\perp \sim 0.01\text { eV}$, one gets $T_D\sim 65 \text{ K}$. On the other hand, below an even lower temperature, say $T_Q$, quantum corrections, not considered in this work, will become important. The predicted effect will, therefore, be observable for $T_Q\lesssim T\lesssim T_D$, and is illustrated in terms of resistivity in Fig.~\ref{resistivity}.

%Discussion% 

\emph{Discussion.} The expression for the surface conductivity used in Eq.~(\ref{dccon}) considers electrons restricted to move along the top layer. Nevertheless, electrons on the surface can tunnel to other layers and reappear on the surface. Such processes also contribute to the surface conductivity, but have been ignored so far. Including these effects, however, does not change the qualitative results. As shown in SM \cite{SM}, the $\tau$- and $T$-dependences of $\sn$ and $\sa$ remain unchanged at low temperature and doping. The only consequence of these extra terms is a renormalization of $W_\mk^2$ in $\sn$: $W_\mk^2 \rightarrow\tilde{W}_\mk^2 = W_\mk^2/(1-W_\mk^2)$ in Eq.~(\ref{sigmaarcint}) while $\sa$ in Eq.~(\ref{sigmabcT}) receives corrections that are suppressed by powers of $|\mu|/t_\perp$ and $k_BT/t_\perp$.

Another aspect of our calculations that need scrutiny is that our starting model in Eq.~(\ref{ham}) defines a WSM which assumes the bilayer basis in real space to be identical to the band basis in energy space. In general, this need not be true. However, in such cases, the bilayer basis and the low energy band basis can be related by a $\mk$-dependent unitary transformation $U_\mk$. In SM \cite{SM} we show that $U_\mk$ does not modify the salient qualitative features $\sigma_\textrm{surf}$. This is essentially because $U_\mk$ does not change the analytic structure of $G_\mk(E)$, which continues to retain poles along the FAs and BCs within the bulk bands. 

Finally, we note that, while $\sigma_\textrm{surf}$ stems from topological features such as the FAs and inseparability of the surface from the bulk, the response itself is non-topological as it does not descend from a topological response theory. Indeed, as shown in SM \cite{SM}, an ordinary metal also acquires a $T^2$ term in the surface conductivity. However, this term vanishes as the bulk Fermi surface shrinks to zero, in contrast to $\sa$ which is non-zero at finite $T$ even when the Weyl nodes are undoped. Additionally, the $\mu$ and $T$ dependences in ordinary metals are essentially unrelated whereas $\sa$ only depends on the combination $\mu^2+k_B^2T^2\pi^2/3$ according to (\ref{sigmabcT}). Both these properties stem from the innate linear dispersion of Weyl fermions. In this sense, $\sigma_\textrm{surf}$ is the surface counterpart of the bulk longitudinal conductivity that stems from topological Weyl fermions but is itself non-topological \cite{WeylMultiLayer,Hosur2012}. Nevertheless, it paves the way to search for other -- possibly topological -- surface responses to complement the extensively studied bulk topological responses originating from the Weyl nodes.

\emph{Conclusion.} We have shown that the dc surface conductivity of a WSM contains an anomalous contribution at nonzero temperature and doping in addition to the expected Drude contribution from the FAs. The novel contribution is manifestly non-Drude in character, being independent of the in-plane scattering time, stems from the intrinsic inseparability of the surface from the bulk, and dominates the temperature dependence at low temperatures. Moreover, its temperature and doping dependences are locked to each other by a universal, dimensionless ratio: $k_B^2\pi^2/3$. We argued that the bulk states when projected on the surface mimics a correlated liquid, and the anomalous conductivity can be understood as a response of this liquid.

%Acknowledgments%

\begin{acknowledgments}
H. K. P. would like to thank IRCC, IIT Bombay for financial support via grant RD/0518-IRCCSH0-029. O. E. O. and P. H. would like to acknowledge financial support from the National Science Foundation grant no. DMR-2047193.
\end{acknowledgments}

%%%%%%%%%%%%%%%%%%%%%%%%%%%%%%%%%%%%%%

\bibliography{paper.bbl}

%%%%%%%%%%%%%%%%%%%%%%%%%%%%%%%%%%%%%

%%%%supplemental%%%%

\begin{widetext}

\section{Supplemental Material}

\section{Surface Green's function\label{surfgreen}}

The Hamiltonian for a slab of a Weyl semimetal (WSM) used in the main text is
\beq
H_{\mk}=\sum_{z=1}^L\psi_{z,\mk}^{\dagger}[(-1)^z\xi_{\mk}-\mu]\psi_{z,\mk}-\sum_{z=1}^{L-1}\psi_{z,\mk}^{\dagger}t_{z,\mk}\psi_{z+1,\mk}+\mathrm{H.c.},
\label{ham}
\eeq
where $\psi_{z,\mk}^\dagger$ creates an electron with 2D momentum $\hbar\mk$ in layer $z$, $\xi_{\mk}=\ve_\mk-\ve_F^{2D}$ denotes a metallic dispersion in each layer with a Fermi surface along $\xi_\mk = 0$, $\mu$ is the bulk chemical potential and the interlayer hopping modulates as $t_{z,\mk}=t_\perp +(-1)^z\delta t_\mk$. In the bulk, it results in the 3D Bloch Hamiltonian and dispersion:
\begin{eqnarray}
H_\mk(k_z) &=& -\mu+\xik\sigma_z - \left[t_\perp-\delta\tk +(t_\perp+\delta\tk)\cos(2k_zc)\right]\sigma_x + t_\perp\sin(2k_zc)\sigma_y\\
\varepsilon_\mk^\pm(k_z) &=& -\mu \pm \sqrt{\xik^2+\left[t_\perp-\delta\tk +(t_\perp+\delta\tk)\cos(2k_zc)\right]^2+t_\perp^2\sin^2(2k_zc)}
\end{eqnarray}
The Green's function $\hat{G}_L(E,\mk)$ is found by computing $(E-H_\mk)^{-1}$. Written explicitly in the layer basis,
\beq
\hat{G}_L(E,\mk)=
\begin{pmatrix}
E-\xi_\mk-\mu & t_{\perp}-\delta t_\mk & & & & &\\
t_{\perp}-\delta t_\mk & E+\xi_\mk-\mu & t_{\perp}+\delta t_\mk & & & & \\
&t_{\perp}+\delta t_\mk& E-\xi_\mk-\mu & t_{\perp}-\delta t_\mk & & &\\
& & t_{\perp}-\delta t_\mk & E+\xi_\mk-\mu & t_{\perp}+\delta t_\mk & &\\
& & & & \ddots & &
\end{pmatrix}^{-1}.
\label{ggg}
\eeq
The above can be written in the form
\begin{equation}
\hat{G}_L(E,\mk)=\left(\begin{array}{cc}
X & Y\\
Y^{\dagger} & Z
\end{array}\right)^{-1},
\end{equation}
where 
\beq
X=
\begin{pmatrix}
E-\xi_\mk-\mu & t_{\perp}-\delta t_\mk \\
t_{\perp}-\delta t_\mk & E+\xi_\mk-\mu  
\end{pmatrix};\ \ 
Y=
\begin{pmatrix}
0 & 0 & 0 & \cdots \\
t_{\perp}-\delta t_\mk & 0 & 0 & \cdots 
\end{pmatrix};\ \
Z=
\begin{pmatrix}
E-\xi_\mk-\mu & t_{\perp}-\delta t_\mk & & & & &\\
t_{\perp}-\delta t_\mk & E+\xi_\mk-\mu & t_{\perp}+\delta t_\mk & & & & \\
&t_{\perp}+\delta t_\mk& \ddots & & & &
\end{pmatrix} .
\eeq
We now use the identity for block matrix inversion,
\begin{equation}
\left(\begin{array}{cc}
X & Y\\
Y^{\dagger} & Z
\end{array}\right)^{-1}=\left(\begin{array}{cc}
(X-YZ^{-1}Y^{\dagger})^{-1} & -(X-YZ^{-1}Y^{\dagger})^{-1}YZ^{-1}\\
-Z^{-1}Y^{\dagger}(X-YZ^{-1}Y^{\dagger})^{-1} & Z^{-1}+Z^{-1}Y^{\dagger}(X-YZ^{-1}Y^{\dagger})^{-1}YZ^{-1}
\end{array}\right)\label{eq:block-matrix-inversion}
\end{equation}
to calculate Eq.~(\ref{ggg}). Since we are interested in the surface Green's function $G_\mk(E)\equiv \hat{G}_L^{11}(E,\mk)$, Eq. (\ref{eq:block-matrix-inversion}) implies
\beq
\hat{G}_L^{11}(E,\mk)=[(X-YZ^{-1}Y^{\dagger})^{-1}]^{11}.
\label{gg11}
\eeq
But $Z^{-1}\equiv \hat{G}_{L-2}(E,\mk)$. Using this in Eq.~(\ref{gg11}), and approximating $\hat{G}_L^{11}(E,\mk)=\hat{G}_{L-2}^{11}(E,\mk)$ as $L\rightarrow\infty$, we find the expression for surface Green's function for a semi-infinite Weyl semimetal:
\begin{eqnarray}
G_{\mk}(E)&=&\frac{a_{\mk}(E)+b_{\mk}(E)}{2(t_\perp+\delta\tk)^2(E+\mu-\xik)},\label{G11}\\
a_{\mk}(E)&=&(E+\mu)^2-\xik^2+4t_\perp\delta t_\mk,\nonumber\\
b_{\mk}(E)&=&\sqrt{\left[(E+\mu)^2-\xik^2-4t_\perp^2\right]\left[(E+\mu)^2-\xik^2-4\delta t_\mk^2\right]}.\nonumber
\end{eqnarray}
This expression, in the Matsubara representation, has been quoted in the main text. 

The above surface Green's function is derived for a clean system. For transport, we need to find the equivalent expression in the presence of scattering. To incorporate such effects, we take a phenomenological route without getting into the microscopics. In the presence of scattering, energy eigenstates in each layer will gain a self-energy term. Since there are two kinds of layers, electron-like and hole-like, we introduce two kinds of self-energies. This is equivalent to making the following substitution in Eq.~(\ref{ham}): $(-1)^z\xi_\mk\rightarrow (-1)^z\xi_\mk+\Sigma_z$, with $\Sigma_z=\Sigma_{e(h)}$ for $z$ even(odd). Going through the same steps as before, we find
\begin{eqnarray}
G_{\mk}(E)&=&\frac{a_{\mk}(E)+b_{\mk}(E)}{2(t_\perp+\delta\tk)^2(E+\mu-\xik+\Sigma_e)},\label{G11}\\
a_{\mk}(E)&=&\left(E+\mu+\frac{\Sigma_e+\Sigma_h}{2}\right)^2-\left(\xik-\frac{\Sigma_e-\Sigma_h}{2}\right)^2+4t_\perp\delta t_\mk,\nonumber\\
b_{\mk}(E)&=&\left\{\left[\left(E+\mu+\frac{\Sigma_e+\Sigma_h}{2}\right)^2-\left(\xik-\frac{\Sigma_e-\Sigma_h}{2}\right)^2-4t_\perp^2\right]\right.\nonumber\\
&\times&\left.\left[\left(E+\mu+\frac{\Sigma_e+\Sigma_h}{2}\right)^2-\left(\xik-\frac{\Sigma_e-\Sigma_h}{2}\right)^2-4\delta t_\mk^2\right]\right\}^{1/2}.\nonumber
\end{eqnarray}
In the presence of weak scattering, all self-energy terms in the numerator of Eq.~(\ref{G11}) can be neglected leaving only the one in the denominator. Rewriting $\mr{Im}[\Sigma_e]$ as $(i\hbar/2\tau_\mk) \mr{sgn}(E+\mu)$ and analytically continuing, one can get the retarded Green's functions as usual. However, note that the Green's function has branch cuts (BCs) in addition to poles: the former occur when $|E+\mu|\in[E_{\mk-},E_{\mk+}]$ -- the BC region -- and the latter occur when $|E+\mu|\not\in[E_{\mk-},E_{\mk+}]$ -- the normal (N) region, where $E_{\mk+}=\sqrt{\xik^2+4t_\perp^2}=\varepsilon_\mk(0)+\mu$ and $E_{\mk-}=\sqrt{\xik^2+4\delta t_\mk^2}=\varepsilon_\mk(\pi/2c)+\mu$. Accordingly, analytic continuation gives rise to two kinds of expressions in the two regions:
\begin{eqnarray}
G^R_{\mk}(E)&=&\frac{a_{\mk}(E)+b_{\mk}(E)}{2(t_\perp+\delta\tk)^2(E+\mu-\xik+ \frac{i\hbar}{2\tau_\mk})}, E\in N,\label{gn}\\
G^R_{\mk}(E)&=&\frac{a_{\mk}(E)+ i\ \mathrm{sgn}(E+\mu)|b_{\mk}(E)|}{2(t_\perp+\delta\tk)^2(E+\mu-\xik+ \frac{i\hbar}{2\tau_\mk})}, E\in BC.\label{gbc}
\end{eqnarray}
These have been quoted as (4) and (5) in the main text.

\section{General expression for $\sigma_{\text{surf}}$}

In this section, we derive a general expression for $\sigma_{\text{surf}}$
in terms of spectral functions that includes all possible interlayer
processes and is applicable to arbitrary tight-binding Hamiltonians.
We then rederive the results for $\sigma_{\text{norm}}$ and $\sigma_{\text{anom}}$
presented in the main paper in Sec. \ref{sec:Derivation} and analyze
corrections to these results in Sec. \ref{sec:Corrections}.

According to the Kubo formula, the surface ac conductivity as a function
of Matsubara frequency is given by 
\begin{equation}
\sigma(i\omega_{n})=e^{2}\frac{T}{\omega_{n}}\sum_{iE_{n}}\intop_{\mathbf{k}}\left[\hat{v}_{\mathbf{k},x}\hat{G}_{\mathbf{k}}(iE_{n})\hat{v}_{\mathbf{k},x}\hat{G}_{\mathbf{k}}(iE_{n}+i\hbar\omega_{n})\right]_{11}\label{eq:sigma(iw)}
\end{equation}
where $\hat{v}_{\mathbf{k},x}$ and $\hat{G}_{\mathbf{k}}(iE_{n})$\textbf{
}are matrices in layer space, $\left[\dots\right]_{11}$ denotes the
$(1,1)$ matrix element and the surface is defined to be the $z=1$
layer. Evaluating the Matsubara sum via appropriate contour integrals
and analytically continuing to real frequencies, $i\omega_{n}\to\omega+i0^{+}$,
gives the real part of the conductivity
\begin{equation}
\text{Re}\sigma(\omega)=\frac{e^{2}\hbar}{4}\intop_{E,\mathbf{k}}\frac{f(E-\hbar\omega/2)-f(E+\hbar\omega/2)}{\omega}\left[\hat{v}_{\mathbf{k},x}\hat{A}_{\mathbf{k}}(E-\hbar\omega/2)\hat{v}_{\mathbf{k},x}\hat{A}_{\mathbf{k}}(E+\hbar\omega/2)+\textrm{h.c.}\right]_{11}
\end{equation}
where $\hat{A}_{\mathbf{k}}(E)=-i\left[\hat{G}_{\mathbf{k}}(E+i0^{+})-\hat{G}_{\mathbf{k}}^{\dagger}(E+i0^{+})\right]$
is the spectral function matrix in layer space and we have dropped
a term $\propto\delta(\omega)$ that implies a static limit. In the
physical dc limit, $\omega\to0$, we get 
\begin{equation}
\sigma_{\text{surf}}^{\text{tot}}=\text{Re}\sigma(0)=\frac{e^{2}\hbar}{4}\intop_{E,\mathbf{k}}\left(-\frac{\partial f}{\partial E}\right)\left[\hat{v}_{\mathbf{k},x}\hat{A}_{\mathbf{k}}(E)\hat{v}_{\mathbf{k},x}\hat{A}_{\mathbf{k}}(E)+\textrm{h.c.}\right]_{11}\label{eq:sigma_general}
\end{equation}
The factor $-\partial_{E}f$ above restricts the integrand to the
regime $|E|\lesssim T$; the contributions from larger $|E|$ decay
exponentially since $-\partial_{E}f\approx T^{-1}e^{-|E|/T}\text{ for }|E|\gtrsim T$.
Physically, this is the statement that longitudinal dc transport on
WSM surfaces, just like in ordinary metals, is dominated by bands
near the Fermi level with exponentially small corrections from higher
bands. In the rest of the discussion, we will assume $|E|\lesssim T$.

\section{Derivation of main results\label{sec:Derivation}}

Now, we focus on the layering model described in Eq. (2) and restrict
to the $(1,1)$ matrix elements of both $\hat{v}_{\mathbf{k},x}$
and $\hat{A}_{\mathbf{k}}(E)$. In the main paper, we denoted $v_{\mathbf{k},x}^{11}=\frac{1}{\hbar}\partial_{k_{x}}\xi_{\mathbf{k}}$
and $A_{\mathbf{k}}^{11}(E)$ as simply $v_{\mathbf{k},x}$ and $A_{\mathbf{k}}(E)$,
respectively. Explicitly,
\begin{align}
A_{\mathbf{k}}(E) & =2\pi\delta_{\hbar/2\tau_{\mathbf{k}}}(E+\mu-\xi_{\mathbf{k}})W_{\mathbf{k}}+\frac{\text{sgn}(E+\mu)\sqrt{R\left[\left((E+\mu)^{2}-E_{\mathbf{k}-}^{2}\right)\left(E_{\mathbf{k}+}^{2}-(E+\mu)^{2}\right)\right]}}{(t_{\perp}+\delta t_{\mathbf{k}})^{2}(E+\mu-\xi_{\mathbf{k}})}\nonumber \\
 & \equiv A_{\text{norm},\mathbf{k}}(E)+A_{\text{anom},\mathbf{k}}(E)\label{eq:A-parts}
\end{align}
in the limit of small $\hbar/2\tau_{\mathbf{k}}$, where $\delta_{\hbar/2\tau}(E)=\frac{1}{\pi}\frac{\hbar/2\tau}{E^{2}+(\hbar/2\tau)^{2}}$.
Clearly, $A_{\text{norm},\mathbf{k}}(E)$ captures quasiparticle poles
at the Fermi arc and is non-vanishing only in this region, while $A_{\text{anom},\mathbf{k}}(E)$
is non-zero only when $E\in BC$. These matrix elements yield the
contributions to $\sigma_{\text{surf}}^{\text{tot}}$ from processes
on the top layer only, which we denoted $\sigma_{\text{surf}}$ in
the main paper. $\sigma_{\text{surf}}$ then decomposes as
\begin{align}
\sigma_{\text{surf}} & =\frac{e^{2}\hbar}{2}\intop_{E,\mathbf{k}}\left(-\frac{\partial f}{\partial E}\right)v_{\mathbf{k},x}^{2}\left[A_{\text{norm},\mathbf{k}}^{2}(E)+2A_{\text{norm},\mathbf{k}}(E)A_{\text{anom},\mathbf{k}}(E)+A_{\text{anom},\mathbf{k}}^{2}(E)\right]\nonumber \\
 & \equiv\sigma_{\text{norm}}+\sigma_{\text{cross}}+\sigma_{\text{anom}}\label{eq:norm+cross+anom}
\end{align}
Since the pole and one branch point coincide precisely at the Weyl
nodes, but the branch cuts are separated in energy from the poles
everywhere else, $\sigma_{\text{cross}}=e^{2}\hbar\intop_{E,\mathbf{k}}\left(-\frac{\partial f}{\partial E}\right)v_{\mathbf{k},x}^{2}A_{\text{norm},\mathbf{k}}(E)A_{\text{anom},\mathbf{k}}(E)$
vanishes when $1/\tau_{\mathbf{k}}=0$. For non-zero $1/\tau_{\mathbf{k}}$,
the leading $\tau$-dependence is determined by the broadening of
the pole into the branch cut region and is expected to be subleading,
$\leq O(1/\tau)$. Thus, neglecting $\sigma_{\text{cross}}$, $\sigma_{\text{surf}}$
decomposes into $\sigma_{\text{norm}}$ and $\sigma_{\text{anom}}$
as stated in Eq. (1). There, we used an alternate argument for such
a decomposition, namely, by splitting the energy integral as $\intop_{E}=\intop_{N}+\intop_{BC}$.

\subsection{$\sigma_{\text{norm}}$ }

The explicit expression for $\sigma_{\text{norm}}$ according to (\ref{eq:norm+cross+anom})
is
\begin{equation}
\sigma_{\text{norm}}=\frac{e^{2}\hbar}{2}\intop_{E,\mathbf{k}}\left(-\frac{\partial f}{\partial E}\right)v_{\mathbf{k},x}^{2}A_{\text{norm},\mathbf{k}}^{2}(E)
\end{equation}
Writing $A_{\text{norm},\mathbf{k}}^{2}(E)=2\pi W_{\mathbf{k}}^{2}\delta(E+\mu-\xi_{\mathbf{k}})\times2\pi\delta_{\hbar/2\tau}(0)=4\pi W_{\mathbf{k}}^{2}\tau_{\mathbf{k}}\delta(E+\mu-\xi_{\mathbf{k}})/\hbar$,
we recover Eq. (7).

\subsection{$\sigma_{\text{anom}}$\label{subsec:sigma_anom}}

From (\ref{eq:norm+cross+anom}), $\sigma_{\text{anom}}$ is given
by 
\begin{align}
\sigma_{\text{anom}} & =\frac{e^{2}\hbar}{2}\intop_{E,\mathbf{k}}\left(-\frac{\partial f}{\partial E}\right)v_{\mathbf{k},x}^{2}A_{\text{anom},\mathbf{k}}^{2}(E)\\
 & =\frac{e^{2}\hbar}{2}\intop_{E\in BC}\intop_{\mathbf{k}}\left(-\frac{\partial f}{\partial E}\right)v_{\mathbf{k},x}^{2}\frac{\left((E+\mu)^{2}-E_{\mathbf{k}-}^{2}\right)\left(E_{\mathbf{k}+}^{2}-(E+\mu)^{2}\right)}{(t_{\perp}+\delta t_{\mathbf{k}})^{4}(E+\mu-\xi_{\mathbf{k}})^{2}}\label{eq:sigma-BC-general}
\end{align}
As argued in the main paper, the combination $\intop_{E\in BC}\partial_{E}f$
is negligible far from the Weyl nodes. Linearizing around the nodes
as 
\begin{equation}
\xi_{\mathbf{k}}\approx\hbar\mathbf{v}_{j}\cdot\mathbf{q}_{2D}\,,\,\delta t_{\mathbf{k}}\approx\frac{\hbar}{2}\mathbf{u}_{j}\cdot\mathbf{q}_{2D}\label{eq:linearize}
\end{equation}
and reversing the order of the $E$ and momentum integrals gives
\begin{equation}
\sigma_{\text{anom}}=2e^{2}\hbar\sum_{j}\frac{v_{j,x}^{2}}{t_{\perp}^{2}}\intop_{E}\left(-\frac{\partial f}{\partial E}\right)\intop_{\hbar\sqrt{(\mathbf{v}_{j}\cdot\mathbf{q}_{2D})^{2}+(\mathbf{u}_{j}\cdot\mathbf{q}_{2D})^{2}}<|E+\mu|}\frac{d^{2}q}{(2\pi)^{2}}\frac{(E+\mu)^{2}-\hbar^{2}(\mathbf{v}_{j}\cdot\mathbf{q}_{2D})^{2}-\hbar^{2}(\mathbf{u}_{j}\cdot\mathbf{q}_{2D})^{2}}{(E+\mu-\hbar\mathbf{v}_{j}\cdot\mathbf{q}_{2D})^{2}}
\end{equation}
Note, $\mathbf{v}_{j}$ and $\mathbf{u}_{j}$ are not necessarily
orthogonal. To simplify the above integrals, we define $q_{\parallel}$
and $q_{\perp}$ as the components of $\mathbf{q}_{2D}$ parallel
and perpendicular to $\mathbf{v}_{j}$, respectively, and perform
suitable rotation and scaling transformations:
\begin{equation}
\left(\begin{array}{c}
Q\cos\phi\\
Q\sin\phi
\end{array}\right)=\left(\begin{array}{cc}
V_{j} & 0\\
0 & U_{j}
\end{array}\right)\left(\begin{array}{cc}
\cos\theta_{j} & \sin\theta_{j}\\
-\sin\theta_{j} & \cos\theta_{j}
\end{array}\right)\left(\begin{array}{c}
q_{\parallel}\\
q_{\perp}
\end{array}\right)\label{eq:transformations}
\end{equation}
so that $(\mathbf{v}_{j}\cdot\mathbf{q}_{2D})^{2}+(\mathbf{u}_{j}\cdot\mathbf{q}_{2D})^{2}=Q^{2}$.
These transformations imply the following relations that we will need
shortly: 
\begin{align}
\implies & \frac{1}{v_{j}}=\left|\frac{\cos\theta_{j}}{V_{j}}+i\frac{\sin\theta_{j}}{U_{j}}\right|\,,\,U_{j}V_{j}=|\mathbf{v}_{j}\times\mathbf{u}_{j}|
\end{align}
We now get
\begin{align}
\sigma_{\text{anom}} & =2e^{2}\hbar\sum_{j}\frac{v_{j,x}^{2}}{t_{\perp}^{2}}\intop_{E}\left(-\frac{\partial f}{\partial E}\right)\intop_{\hbar Q<|E+\mu|}\frac{d^{2}Q}{(2\pi)^{2}U_{j}V_{j}}\frac{(E+\mu)^{2}-\hbar^{2}Q^{2}}{[E+\mu-\hbar v_{j}Q(\cos\phi\cos\theta_{j}/V_{j}+\sin\phi\sin\theta_{j}/U_{j})]^{2}}
\end{align}
The $\phi$-integral is most easily performed via contour methods.
Defining $\zeta=e^{i\phi}$, $\beta=(E+\mu)/\hbar Q$ and $e^{i\alpha}=v_{j}\left(\frac{\cos\theta_{j}}{V_{j}}-i\frac{\sin\theta_{j}}{U_{j}}\right)$,
\begin{align}
\sigma_{\text{anom}} & =2e^{2}\hbar\sum_{j}\frac{v_{j,x}^{2}}{t_{\perp}^{2}}\intop_{E}\left(-\frac{\partial f}{\partial E}\right)\intop_{\hbar Q<|E+\mu|}\frac{QdQ}{(2\pi)^{2}iU_{j}V_{j}}\oint\frac{4(\beta^{2}-1)\zeta d\zeta}{\left(2\beta\zeta-\left[\zeta^{2}e^{i\alpha}+e^{-i\alpha}\right]\right)^{2}}
\end{align}
The poles of the $\zeta$-integrand in both expressions are at
\begin{equation}
\zeta_{\pm}=e^{-i\alpha}\left(\beta\pm\sqrt{\beta^{2}-1}\right)
\end{equation}
of which exactly one is within the unit circle in the complex plane.
Integrating over the unit circle now gives

\begin{align}
\sigma_{\text{anom}} & =e^{2}\hbar\sum_{j}\frac{v_{j,x}^{2}}{t_{\perp}^{2}}\intop_{-\infty}^{\infty}\frac{dE}{2\pi}\left(-\frac{\partial f}{\partial E}\right)\intop_{\hbar Q<|E+\mu|}\frac{QdQ}{2\pi U_{j}V_{j}}\frac{2|\beta|}{\sqrt{\beta^{2}-1}}
\end{align}
The $Q$ and $E$ integrals are trivial, and yield 
\begin{equation}
\sigma_{\text{anom}}=\frac{e^{2}}{\hbar\pi^{2}}\sum_{j}\frac{v_{j,x}^{2}}{2t_{\perp}^{2}|\mathbf{u}_{j}\times\mathbf{v}_{j}|}\left(\mu^{2}+\frac{T^{2}\pi^{2}}{3}\right)
\end{equation}
where we have used a Sommerfeld expansion for $f(E)$ and the relation
$U_{j}V_{j}=|\mathbf{u}_{j}\times\mathbf{v}_{j}|$. Recalling that
$\mathbf{w}_{j}=-\frac{2}{\hbar}t_{\perp}c\hat{\mathbf{z}}$ gives
Eq. (10).

\section{Corrections to $\sigma_{\textrm{surf}}$\label{sec:Corrections}}

In this section, we analyze corrections to the results for $\sigma_{\text{norm}}$
and $\sigma_{\text{anom}}$ in the main paper. To compute the corrections,
we will need some other matrix elements $\hat{G}_{\mathbf{k}}(E)$,
which we first derive.

\subsection{Other matrix elements of $\hat{G}_{\mathbf{k}}(E)$}

Consider adding $p$ layers to a general $N$-layered system. The
Hamiltonian for the $N+p$-layered system at fixed transverse momentum
may be written as
\begin{equation}
H_{N+p}^{0}=\left(\begin{array}{ccccc}
H_{p}^{0} & H_{p}^{-} & 0 & 0 & \dots\\
H_{p}^{+}\\
0 &  & H_{N}^{0}\\
0\\
\vdots
\end{array}\right)
\end{equation}
where $H_{p}^{0,\pm}$ are $p\times p$ matrices, $H_{N}^{0}$ is
$N\times N$ and we have suppressed $\mathbf{k}$-dependence for clarity.
Note that $H_{p}^{-}$ is lower triangular and $H_{p}^{+}=(H_{p}^{-})^{\dagger}$.
We define $\hat{G}_{N}(E)=(E-H_{N}^{0})^{-1}$ and $\hat{G}_{N,p}(E)$
as the top-left $p\times p$ block of $\hat{G}_{N}(E)$ and use the
identity for block matrix inversion
\begin{equation}
\left(\begin{array}{cc}
X & Y\\
Y^{\dagger} & Z
\end{array}\right)^{-1}=\left(\begin{array}{cc}
(X-YZ^{-1}Y^{\dagger})^{-1} & -(X-YZ^{-1}Y^{\dagger})^{-1}YZ^{-1}\\
-Z^{-1}Y^{\dagger}(X-YZ^{-1}Y^{\dagger})^{-1} & Z^{-1}+Z^{-1}Y^{\dagger}(X-YZ^{-1}Y^{\dagger})^{-1}YZ^{-1}
\end{array}\right)\label{eq:block-matrix-inversion}
\end{equation}
with $X=E-H_{p}^{0}$, $Y=\left(H_{p}^{-},0,0,\dots\right)$, $Z=E-H_{N}^{0}$.
The triangularity of $H_{p}^{\pm}$ ensures that $\hat{G}_{N+p,p}(E)$
depends only on $\hat{G}_{N,p}(E)$, i.e., it does not depend on the
elements of $\hat{G}_{N}(E)$ outside the top-left $p\times p$ block.
In particular,
\begin{equation}
\hat{G}_{N+p,p}(E)=\left(E-H_{p}^{0}-H_{p}^{-}G_{N,p}(E)H_{p}^{+}\right)^{-1}
\end{equation}
For $N\gg p$, we expect $\hat{G}_{N+p,p}(E)\approx\hat{G}_{N,p}(E)$,
which gives the self-consistent equation
\begin{equation}
\hat{G}_{N,p}^{-1}(E)=E-H_{p}^{0}-H_{p}^{-}G_{N,p}(E)H_{p}^{+}\label{eq:G-eq}
\end{equation}

Focusing on the layering model in Eq. (2), choosing $p=2$ (one bilayer
unit cell) and $N=L-2$ gives a $2\times2$ symmetric matrix Green's
function for the surface unit cell, $\mathcal{G}_{\mathbf{k}}(E)$,
where we have reinstated the $\mathbf{k}$ and $E$ dependences. In
the main paper, we have denoted $\mathcal{G}_{\mathbf{k}}^{11}(E)\equiv G_{\mathbf{k}}(E)$.
Some algebra on (\ref{eq:G-eq}) gives the other matrix elements of
$\mathcal{G}_{\mathbf{k}}(E)$ as
\begin{align}
\mathcal{G}_{\mathbf{k}}^{12}(E)\equiv G_{\mathbf{k}}^{12}(E) & =-\frac{(E+\mu)^{2}-\xi_{\mathbf{k}}^{2}-4t_{\perp}^{2}-4\delta t_{\mathbf{k}}^{2}+b_{\mathbf{k}}(E)}{2(t_{\perp}-\delta t_{\mathbf{k}})(t_{\perp}+\delta t_{\mathbf{k}})^{2}}\label{eq:G-12}\\
\mathcal{G}_{\mathbf{k}}^{22}(E)\equiv G_{\mathbf{k}}^{22}(E) & =-\frac{E+\mu-\xi_{\mathbf{k}}}{t_{\perp}-\delta t_{\mathbf{k}}}\mathcal{G}_{\mathbf{k}}^{12}(E)\label{eq:G-22}
\end{align}
In the presence of a non-zero $\tau_{\mathbf{k}}$, we must ideally
replace $E\to E\pm i\hbar/2\tau_{\mathbf{k}}$ for the retarded and
advanced parts of the Green's functions. However, we see that $G_{\mathbf{k}}^{12}(E)$
and $G_{\mathbf{k}}^{22}(E)$ have no poles and we can safely set
$1/\tau_{\mathbf{k}}=0$ to leading order. On the other hand, these
matrix elements develop branch cuts along $|E+\mu|\in(E_{\mathbf{k}-},E_{\mathbf{k}+})$,
just as $G_{\mathbf{k}}(E)$ does, when $b_{\mathbf{k}}^{2}(E)<0$.
In fact, these matrix elements are purely real except on the branch
cuts. The spectral functions matrix elements $A_{\mathbf{k}}^{zz'}(E)=-2\text{Im}\left(G_{\mathbf{k}}^{R}(E)\right)^{zz'}$
are given by
\begin{align}
\mathcal{A}_{\mathbf{k}}^{12}(E)\equiv A_{\mathbf{k}}^{12}(E) & =-\frac{E+\mu-\xi_{\mathbf{k}}}{t_{\perp}-\delta t_{\mathbf{k}}}A_{\text{anom},\mathbf{k}}\label{eq:A-12}\\
\mathcal{A}_{\mathbf{k}}^{22}(E)\equiv A_{\mathbf{k}}^{22}(E) & =\left(\frac{E+\mu-\xi_{\mathbf{k}}}{t_{\perp}-\delta t_{\mathbf{k}}}\right)^{2}A_{\text{anom},\mathbf{k}}\label{eq:A-22}
\end{align}
Assuming $|E|\lesssim T$, $(T,|\mu|)\ll|t_{\perp}|$, and the absence
of fine-tuned regions away from the Weyl nodes with $|\delta t_{\mathbf{k}}|\lesssim(T,|\mu|)$,
this means
\begin{equation}
\left|\mathcal{A}_{\mathbf{k}}^{22}(E)\right|\ll\left|\mathcal{A}_{\mathbf{k}}^{12}(E)\right|\ll\left|A_{\text{anom},\mathbf{k}}(E)\right|\ll\left|A_{\text{norm},\mathbf{k}}(E)\right|\label{eq:A-limits}
\end{equation}

The off diagonal block of (\ref{eq:block-matrix-inversion}) yields
a recursion relation for the off-diagonal matrix elements of $\hat{G}$
in the first two rows and columns: 
\begin{align}
G_{\mathbf{k}}^{1z}(E) & =G_{\mathbf{k}}^{z1}(E)=\mathcal{G}_{\mathbf{k}}^{12}(E)(t_{\perp}+\delta t_{\mathbf{k}})G_{\mathbf{k}}^{1,z-2}(E)\,;\,z\geq3\\
G_{\mathbf{k}}^{2z}(E) & =G_{\mathbf{k}}^{z2}(E)=\mathcal{G}_{\mathbf{k}}^{22}(E)(t_{\perp}-\delta t_{\mathbf{k}})G_{\mathbf{k}}^{1,z-2}(E)\,;\,z\geq3
\end{align}
The recursion relations yield the closed form expressions 
\begin{align}
G_{\mathbf{k}}^{1z}(E) & =\begin{cases}
\left[\mathcal{G}_{\mathbf{k}}^{12}(E)(t_{\perp}+\delta t_{\mathbf{k}})\right]^{(z-1)/2}G_{\mathbf{k}}(E) & z\in\text{odd}\\
\left[\mathcal{G}_{\mathbf{k}}^{12}(E)(t_{\perp}+\delta t_{\mathbf{k}})\right]^{z/2-1}\mathcal{G}_{\mathbf{k}}^{12}(E) & z\in\text{even}
\end{cases}\label{eq:G-1z}\\
G_{\mathbf{k}}^{2z}(E) & =-\frac{E+\mu-\xi_{\mathbf{k}}}{t_{\perp}-\delta t_{\mathbf{k}}}G_{\mathbf{k}}^{1z}(E)\,\,z\geq2\label{eq:G-2z}
\end{align}
Now, $G_{\mathbf{k}}^{1z}(E)$ inherits a simple pole from $G_{\mathbf{k}}(E)$
for odd $z\geq3$ but lacks poles for even $z$, while $G_{\mathbf{k}}^{2z}(E)$
lacks poles for any $z$. Using (\ref{eq:G-1z}), (4) and (5), the
corresponding matrix elements of $\hat{A}_{\mathbf{k}}(E)$ are
\begin{align}
A_{\mathbf{k}}^{1z}(E) & =\begin{cases}
-2\pi\left(-W_{\mathbf{k}}\right)^{(z+1)/2}\delta_{\hbar/2\tau_{\mathbf{k}}}(E+\mu-\xi_{\mathbf{k}}) & z\in\text{odd};E\in N\\
0 & z\in\text{even};E\in N\\
-\frac{z+1}{E+\mu-\xi_{\mathbf{k}}}\frac{(2\delta t_{\mathbf{k}})^{(z-1)/2}\text{sgn}(E+\mu)\sqrt{(E+\mu)^{2}-E_{\mathbf{k}-}^{2}}}{(-t_{\perp})^{(z+1)/2}} & z\in\text{odd};E\in BC\\
\frac{z+1}{t_{\perp}}\frac{(2\delta t_{\mathbf{k}})^{z/2-1}\text{sgn}(E+\mu)\sqrt{(E+\mu)^{2}-E_{\mathbf{k}-}^{2}}}{(-t_{\perp})^{z/2}} & z\in\text{even};E\in BC
\end{cases}\label{eq:A-1z}\\
A_{\mathbf{k}}^{2z}(E) & =-\frac{E+\mu-\xi_{\mathbf{k}}}{t_{\perp}-\delta t_{\mathbf{k}}}A_{\mathbf{k}}^{1z}(E)\,\,z=2,3,4\dots;\text{ any }E
\end{align}

\subsection{Corrections due to interlayer processes}

First, we consider corrections due to tunneling to other layers while
working within the layering model. The velocity matrix elements are
given by 
\begin{equation}
v_{\mathbf{k},x}^{zz'}=\frac{(-1)^{z}}{\hbar}\frac{\partial}{\partial k_{x}}\left(\xi_{\mathbf{k}}\delta_{z,z'}-\delta t_{\mathbf{k}}\delta_{z,z'\pm1}\right)\label{eq:v-matrix}
\end{equation}
The two terms in $v_{\mathbf{k},x}^{zz'}$ produce three types of
corrections, depicted in Fig. \ref{fig:Tunneling-processes}. We analyze
these corrections below.

\begin{figure}
\subfloat[]{\includegraphics[width=0.24\columnwidth]{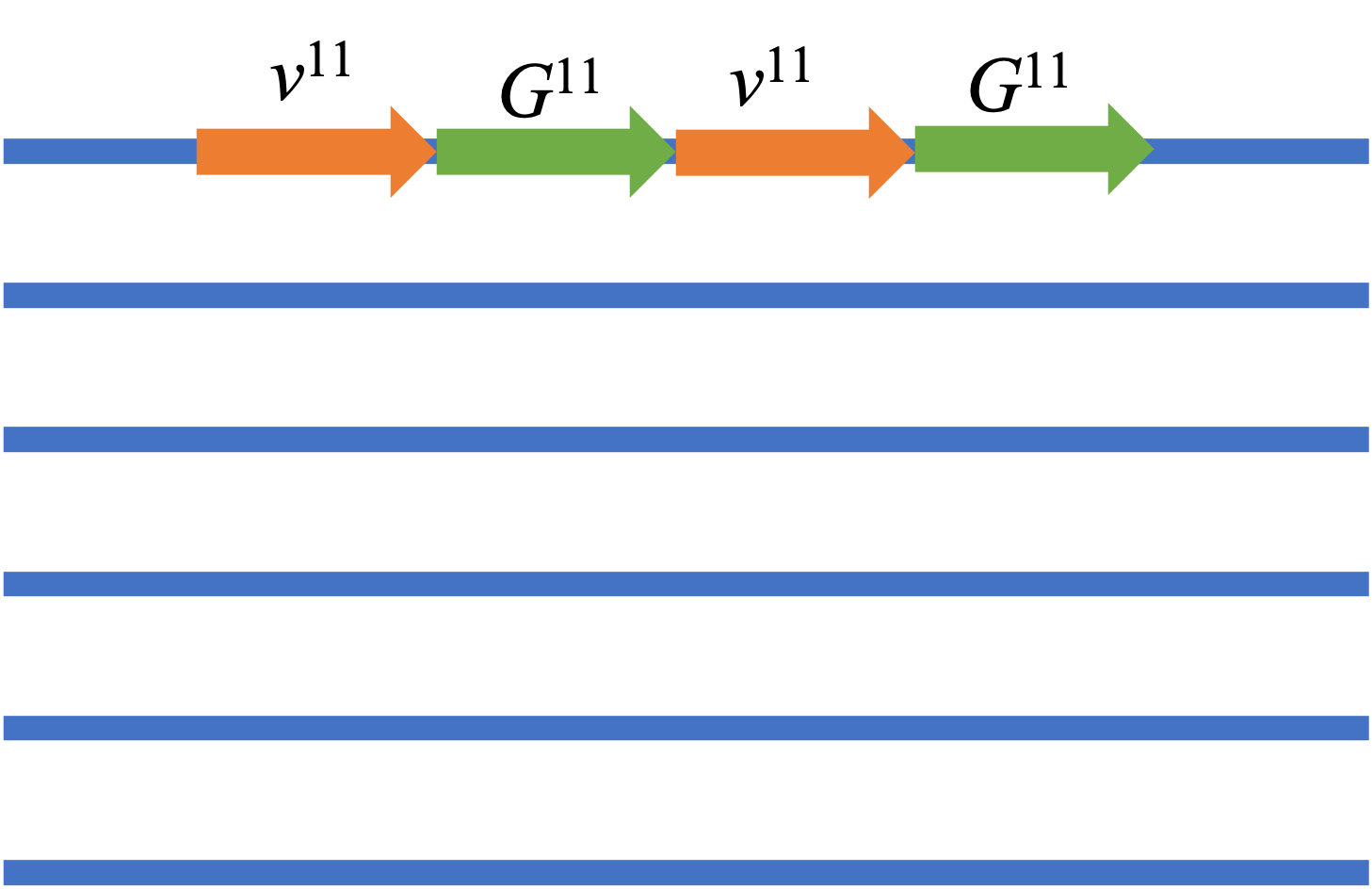}

}~\subfloat[]{\includegraphics[width=0.24\columnwidth]{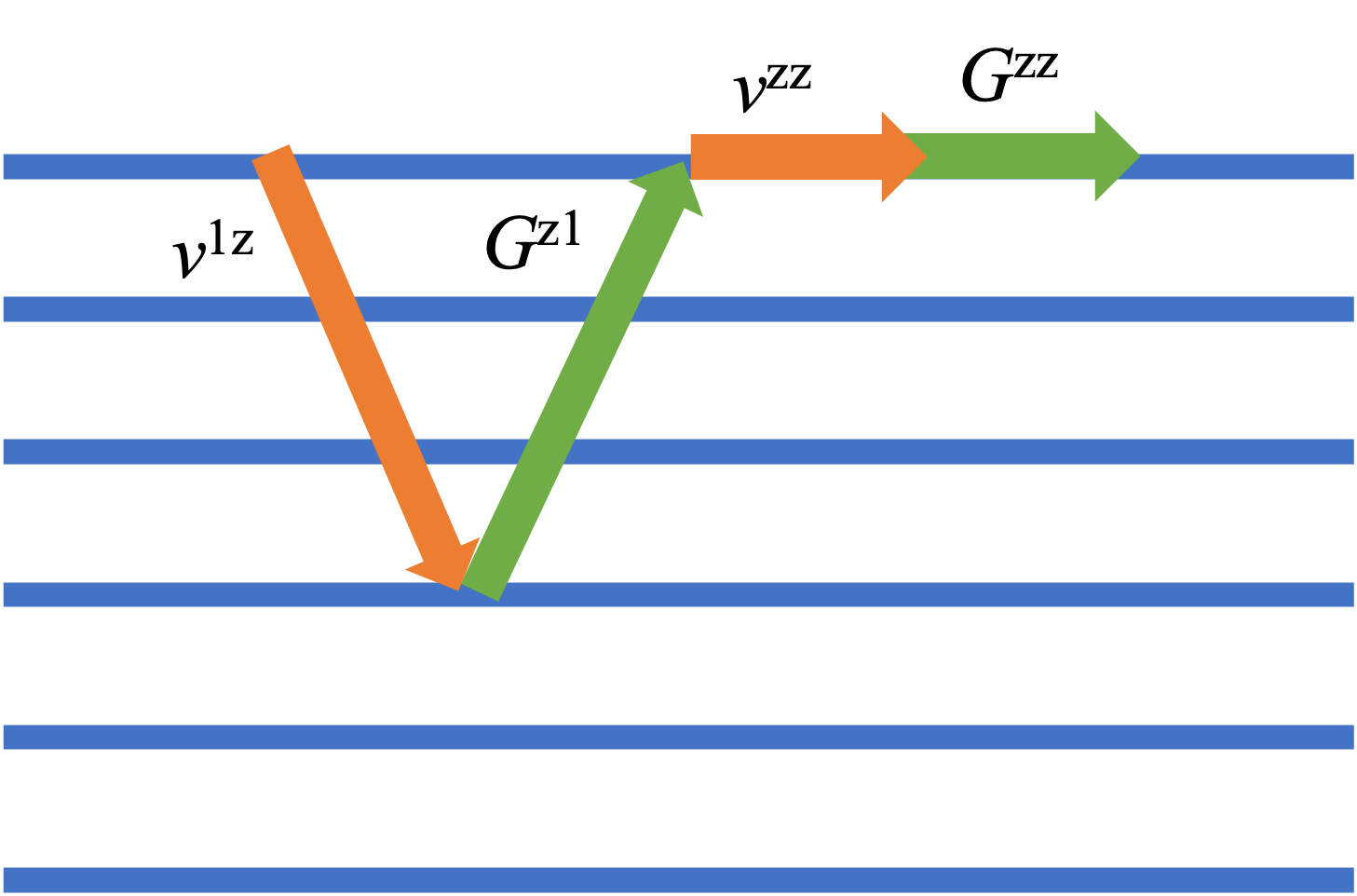}~\includegraphics[width=0.24\columnwidth]{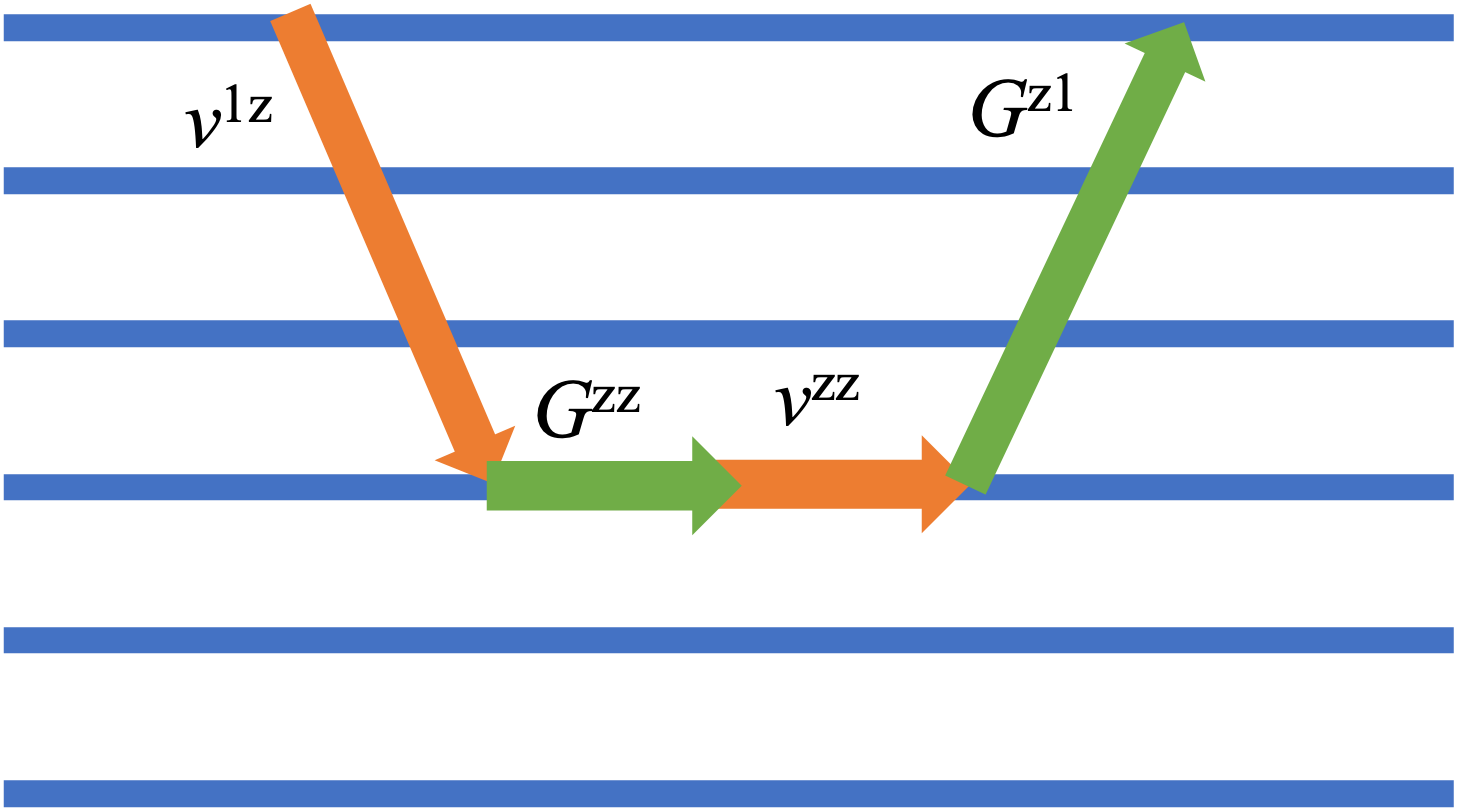}

}~\subfloat[]{\includegraphics[width=0.24\columnwidth]{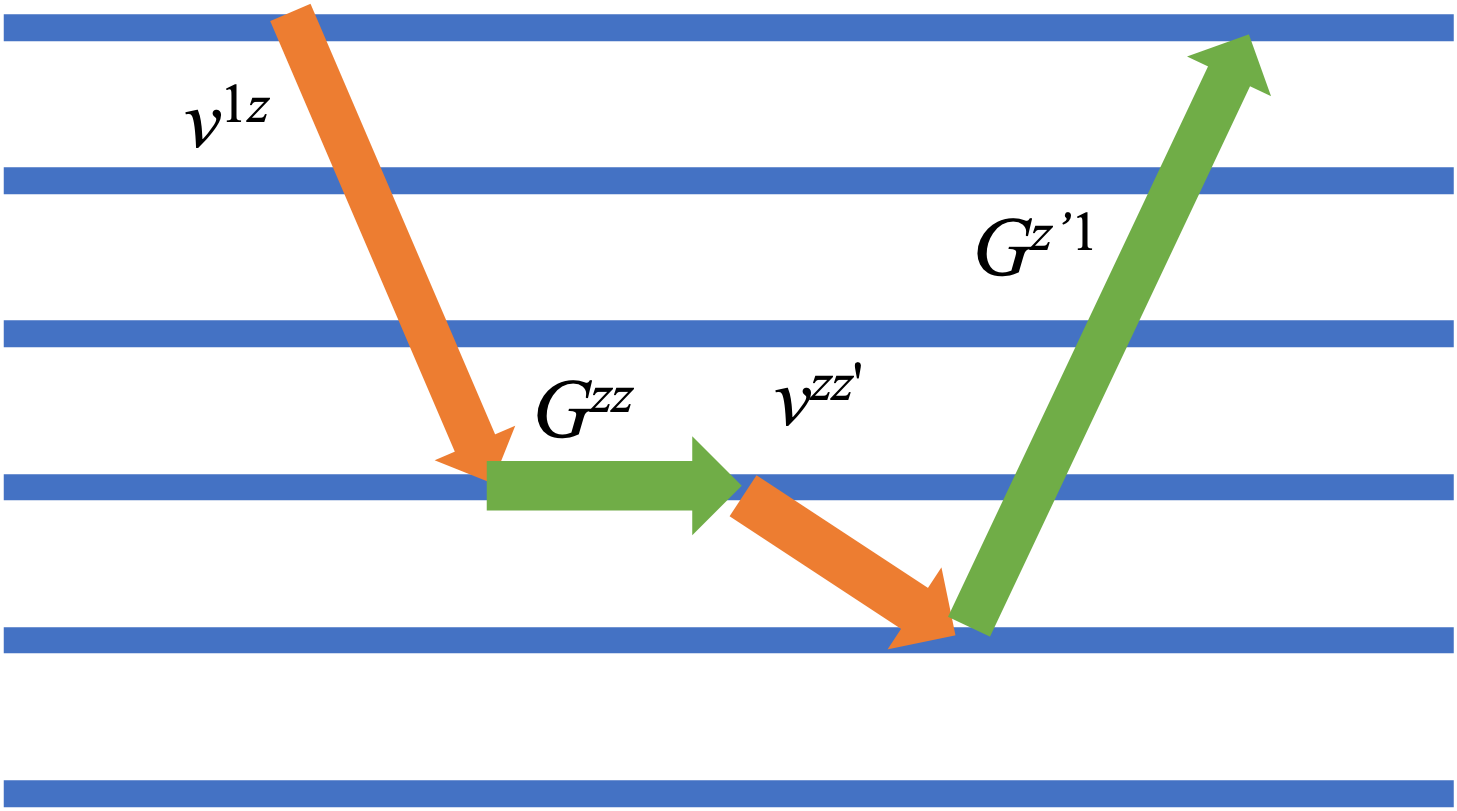}

}

\caption{Schematic of tunneling processes that contribute corrections to the
main result. (a) modifies
$\sigma_{\text{norm}}$ by $\mathcal{O}(1)$, but the modification
can be absorbed into a renormalized weight function $\tilde{W}_{\mathbf{k}}$.
(b) and (c) introduce corrections to $\sigma_{\text{anom}}$, but these are small. See text for details.\label{fig:Tunneling-processes}}
\end{figure}

\subsubsection{$\mathcal{O}\left[\left(\partial_{k_{x}}\xi_{\mathbf{k}}\right)^{2}\right]$
terms in velocity-squared, Fig. \ref{fig:Tunneling-processes}(a)
\label{subsec:vv-terms}}

The corrections due to these terms are 
\begin{equation}
\Delta\sigma_{\text{surf}}^{(a)}=\frac{e^{2}\hbar}{2}\intop_{E,\mathbf{k}}\left(-\frac{\partial f}{\partial E}\right)\sum_{z>1}(-1)^{z+1}\left(v_{\mathbf{k},x}A_{\mathbf{k}}^{1z}(E)\right)^{2}\label{eq:vv-only}
\end{equation}
The first line of (\ref{eq:A-1z}), which corresponds to the Fermi
arc poles, introduces non-negligible corrections into $\sigma_{\text{norm}}$.
These corrections obey a geometric series in $W_{\mathbf{k}}$ and
are easy to evaluate. The result is that $\sigma_{\text{norm}}$ changes
as 
\begin{equation}
\sigma_{\text{norm}}\to\sigma_{\text{norm}}+\Delta\sigma_{\text{norm}}^{(a)}=e^{2}\intop_{\mathbf{k}}2\pi\tau_{\mathbf{k}}\delta(\mu-\xi_{\mathbf{k}})\frac{W_{\mathbf{k}}^{2}}{1-W_{\mathbf{k}}^{2}}v_{\mathbf{k},x}^{2}
\end{equation}
Comparing with Eq. (7), we find that interlayer tunneling renormalizes
the weight factor in $\sigma_{\text{norm}}$ as $W_{\mathbf{k}}^{2}\to W_{\mathbf{k}}^{2}/(1-W_{\mathbf{k}}^{2})$. 

The corrections to $\sigma_{\text{anom}}$, given by the third and
fourth lines of (\ref{eq:A-1z}), are
\begin{equation}
\Delta\sigma_{\text{anom}}^{(a)}=\frac{e^{2}\hbar}{2}\intop_{\mathbf{k}}\intop_{E\in BC}\left(-\frac{\partial f}{\partial E}\right)\sum_{z=2}^{\infty}(-1)^{z+1}\left(v_{\mathbf{k},x}A_{\mathbf{k}}^{1z}(E)\right)^{2}
\end{equation}
Linearizing around the Weyl nodes, integrating over $\mathbf{k}$
using contour methods and Sommerfeld expanding $-f'(E)$ gives
\begin{align}
\Delta\sigma_{\text{anom}}^{(a)} & =\frac{85e^{2}}{94\pi^{2}\hbar}\sum_{j}\frac{v_{j,x}^{2}\mu^{2}\left(\mu^{2}+2T^{2}\pi^{2}\right)}{t_{\perp}^{4}\left|\mathbf{u}_{j}\times\mathbf{v}_{j}\right|}\\
 & =\mathcal{O}\left(\frac{\mu^{2}}{t_{\perp}^{2}}\right)\times\sigma_{\text{anom}}
\end{align}
Thus, the corrections to $\sigma_{\text{anom}}$ due to such interlayer
tunneling processes are negligible. 

\subsubsection{$\mathcal{O}\left[\left(\partial_{k_{x}}\xi_{\mathbf{k}}\right)\left(\partial_{k_{x}}\delta t_{\mathbf{k}}\right)\right]$
terms in velocity squared, Fig. \ref{fig:Tunneling-processes}(b)
\label{subsec:vt-terms}}

The second type of corrections to $\sigma_{\text{surf}}$ come from
including terms that involve interlayer tunneling and one off-diagonal
factor in $\hat{A}_{\mathbf{k}}(E)$. Specifically, the correction
to (\ref{eq:vv-only}) is given by
\begin{align}
\Delta\sigma_{\text{surf}}^{(b)} & =\frac{e^{2}\hbar}{2}\intop_{E,\mathbf{k}}\left(-\frac{\partial f}{\partial E}\right)\sum_{z}v_{\mathbf{k},x}^{12}A_{\mathbf{k}}^{2z}(E)v_{\mathbf{k},x}^{zz}A_{\mathbf{k}}^{z1}(E)
\end{align}
Since $A_{\mathbf{k}}^{2z}(E)\neq0$ only when $E\in BC$ and the
pole and branch cut do not overlap, $\sigma_{\text{norm}}$ receives
no correction from these terms in the limit $1/\tau\to0$:
\begin{equation}
\Delta\sigma_{\text{norm}}^{(b)}=0
\end{equation}
Using (\ref{eq:G-2z}) and (\ref{eq:v-matrix}), the correction to
$\sigma_{\text{anom}}$ is given by
\begin{equation}
\Delta\sigma_{\text{anom}}^{(b)}=\frac{e^{2}}{2}\intop_{\mathbf{k}}\intop_{E\in BC}\left(-\frac{\partial f}{\partial E}\right)\frac{\partial\delta t_{\mathbf{k}}}{\partial k_{x}}v_{\mathbf{k},x}\frac{E+\mu-\xi_{\mathbf{k}}}{t_{\perp}-\delta t_{\mathbf{k}}}\sum_{z}(-1)^{z}\left[A_{\mathbf{k}}^{1z}(E)\right]^{2}
\end{equation}
According to (\ref{eq:A-1z}), $A_{\mathbf{k}}^{1z}(E)\ll A_{\text{anom},\mathbf{k}}(E)$
when $E\in BC$, so the dominant contribution to the above sum is
from $z=1$, which corresponds to Fig. \ref{fig:Tunneling-processes}(b,
left). Retaining only this term, linearizing around the Weyl nodes
and integrating as before gives
\begin{align}
\Delta\sigma_{\text{anom}}^{(b)} & =-\frac{e^{2}}{3\hbar}\sum_{j}\frac{u_{j,x}v_{j,x}}{t_{\perp}^{3}}\frac{|\mu|^{3}+T^{2}\pi^{2}|\mu|}{(2\pi)^{2}\left|\mathbf{u}_{j}\times\mathbf{v}_{j}\right|}\\
 & =\mathcal{O}\left(\frac{|\mu|}{2t_{\perp}}\frac{u_{j,x}}{v_{j,x}}\right)\times\sigma_{\text{anom}}
\end{align}
Clearly, $\sigma_{\text{anom}}$ too receives negligible corrections
at this order in velocity-squared.

\subsubsection{$\mathcal{O}\left[\left(\partial_{k_{x}}\delta t_{\mathbf{k}}\right)^{2}\right]$
terms in velocity squared, Fig. \ref{fig:Tunneling-processes}(c)\label{subsec:tt-terms}}

The correction to $\sigma_{\text{surf}}$ when both velocity factors
come from interlayer tunneling terms is given by
\begin{equation}
\Delta\sigma_{\text{surf}}^{(c)}=\frac{e^{2}\hbar}{2}\intop_{E,\mathbf{k}}\left(-\frac{\partial f}{\partial E}\right)\sum_{z=1}^{\infty}\sum_{z'=z\pm1,z'>0}v_{\mathbf{k},x}^{12}A_{\mathbf{k}}^{2z}(E)v_{\mathbf{k},x}^{zz'}A_{\mathbf{k}}^{z'1}(E)
\end{equation}
The absence of poles in $G_{\mathbf{k}}^{2z}(E)$ according to (\ref{eq:G-2z})
combined with the absence of overlap between the pole and the branch
cut ensures that $E$ is restricted to the BC region in the integral
above, resulting in 
\begin{equation}
\Delta\sigma_{\text{norm}}^{(c)}=0
\end{equation}
Since $A_{\mathbf{k}}^{1z}(E)\ll A_{\text{anom},\mathbf{k}}(E)$ for
$z>1$ and $E\in BC$, the dominant correction to $\sigma_{\text{anom}}$
is given by the $z=2,z'=1$ term. Retaining only this term, Linearizing
around the Weyl nodes and integrating over $\mathbf{k}$ and $E$
as usual yields
\begin{align}
\sigma_{\text{anom}}^{(c)} & =\frac{e^{2}\mu^{2}}{64\pi^{2}\hbar}\left(\mu^{2}+2T^{2}\pi^{2}\right)\sum_{j}\frac{u_{j,x}^{2}}{t_{\perp}^{4}}\frac{1}{U_{j}V_{j}}\\
 & =\mathcal{O}\left(\frac{\mu^{2}}{t_{\perp}^{2}}\frac{u_{j,x}^{2}}{v_{j,x}^{2}}\right)\times\sigma_{\text{anom}}
\end{align}
Thus, the correction due to this term is even smaller.

\subsection{Corrections due to unitary transformations}

The 3D Bloch Hamiltonian generated by Eq. (2) is
\begin{equation}
H_{3D}(\mathbf{k},k_{z})=\xi_{\mathbf{k}}\sigma_{z}-\left[(t_{\perp}+\delta t_{\mathbf{k}})+(t_{\perp}-\delta t_{\mathbf{k}})\cos(2k_{z}c)\right]\sigma_{x}+(t_{\perp}+\delta t_{\mathbf{k}})\sin(2k_{z}c)\sigma_{y}
\end{equation}
where $\sigma_{i}$ are Pauli matrices in the bilayer basis and $\sigma_{z}=\pm1$
correspond to the two layers within a unit cell. At $\delta t_{\mathbf{k}}=0$,
$H_{3D}(\mathbf{k},k_{z})$ realizes a nodal line semimetal with a
line node along $\xi_{\mathbf{k}}=0$ at $k_{z}=\pi/2c$. In the nodal
plane, $k_{z}=2\pi/c$, the Bloch Hamiltonian reduces to a diagonal
form, $H_{3D}(\mathbf{k},k_{z})=\xi_{\mathbf{k}}\sigma_{z}$. In other
words, the two bands that intersect along the line node live on different
layers in the nodal plane, which results in the top layer being purely
hole-like. Note, this conclusion does not require the material to
be layered or quasi-2D; it holds even when $t_{\perp}$ is of the
same order or even larger than the typical scale of $\xi_{\mathbf{k}}$
(denoted $t_{\parallel}$ in the main text).

More generally, the low energy band basis in the bulk would be related
to the bilayer basis by a unitary transformation $U_{\mathbf{k}}$.
Then, the surface conductivity due to motion of electrons on the top
layer is given by 
\begin{equation}
\tilde{\sigma}_{\text{surf}}=\frac{e^{2}\hbar}{2}\intop_{\mathbf{k},E}\left(\tilde{v}_{\mathbf{k},x}^{11}\tilde{A}_{\mathbf{k},x}^{11}\right)^{2}
\end{equation}
where $\tilde{v}_{\mathbf{k},x}^{11}=\frac{1}{\hbar}\partial_{k_{x}}\left(U_{\mathbf{k}}\mathcal{H}_{\mathbf{k}}U_{\mathbf{k}}^{\dagger}\right)_{11}$,
$\tilde{A}_{\mathbf{k}}^{11}(E)=-2\text{Im}\left[U_{\mathbf{k}}\mathcal{G}_{\mathbf{k}}^{R}(E)U_{\mathbf{k}}^{\dagger}\right]_{11}$
and $\mathcal{H}_{\mathbf{k}}=\left(\begin{array}{cc}
-\xi_{\mathbf{k}}-\mu & -t_{\perp}+\delta t_{\mathbf{k}}\\
-t_{\perp}+\delta t_{\mathbf{k}} & \xi_{\mathbf{k}}-\mu
\end{array}\right)$ is the Hamiltonian of the topmost bilayer in Eq. (2). Parameterizing
$U_{\mathbf{k}}=\exp\left(i\boldsymbol{\sigma}\cdot\hat{\mathbf{n}}_{\mathbf{k}}\theta_{\mathbf{k}}\right)$
gives
\begin{align}
\tilde{v}_{\mathbf{k},x}^{11} & =-v_{\mathbf{k},x}\left(\cos2\theta+2\sin^{2}\theta n_{z}^{2}\right)-\frac{\xi_{\mathbf{k}}}{\hbar}\frac{\partial}{\partial k_{x}}\left(\cos2\theta+2\sin^{2}\theta n_{z}^{2}\right)\nonumber \\
 & -\frac{\partial}{\partial k_{x}}\left[(t_{\perp}-\delta t_{\mathbf{k}})\left(n_{y}\sin2\theta+2n_{z}n_{x}\sin^{2}\theta\right)\right]\nonumber \\
 & \equiv-v_{\mathbf{k},x}^{\text{eff}}
\end{align}
and
\begin{align}
\tilde{A}_{\mathbf{k}}^{11}(E) & =\mathcal{A}_{\mathbf{k}}^{11}(E)(\cos\theta_{\mathbf{k}}+n_{\mathbf{k},z}^{2}\sin^{2}\theta_{\mathbf{k}})\nonumber \\
&+\mathcal{A}_{\mathbf{k}}^{22}(E)\sin\theta_{\mathbf{k}}(1-n_{\mathbf{k},z}^{2}\sin\theta_{\mathbf{k}})-\mathcal{A}_{\mathbf{k}}^{12}(E)\left(2n_{\mathbf{k},y}\sin2\theta_{\mathbf{k}}+n_{\mathbf{k},x}n_{\mathbf{k},z}\sin^{2}\theta_{\mathbf{k}}\right)\nonumber \\
 & \approx A_{\mathbf{k}}(E)(\cos\theta_{\mathbf{k}}+n_{\mathbf{k},z}^{2}\sin^{2}\theta_{\mathbf{k}})\\
 & \equiv A_{\mathbf{k}}^{\text{eff}}(E)
\end{align}
using (\ref{eq:A-limits}). While $v_{\mathbf{k},x}^{\text{eff}}$
differs from $v_{\mathbf{k},x}$ by an $O(1)$ factor as well as additive
corrections, $A_{\mathbf{k}}(E)$ merely acquires an $O(1)$ prefactor
due to the unitary transformation. Crucially, the qualitative results,
namely, the decomposition of $\sigma_{\text{surf}}$ into $\sigma_{\text{norm}}$
and $\sigma_{\text{anom}}$ and their $T$, $\tau$ and $\mu$ dependences,
are determined governed by $A_{\mathbf{k}}(E)$ while $v_{\mathbf{k},x}(E)$
only affects the quantitative behavior. Thus, $U_{\mathbf{k}}$ is
expected to leave the qualitative results unchanged.

\section{Comparison with ordinary metals \label{sec:Ordinary-metals}}

In this section, we calculate the surface conductivity of an ordinary metal and compare it with the results obtained for a WSM. Our approach is similar to that used for the WSM. We start with a minimal tight-binding model for an ordinary metal:

\begin{equation}
H_{\text{metal}}=\sum_{z}(\varepsilon_{\mathbf{k}}-\mu)\psi_{z,\mathbf{k}}^{\dagger}\psi_{z,\mathbf{k}}-\sum_{z}t_{\perp}\left(\psi_{z,\mathbf{k}}^{\dagger}\psi_{z+1,\mathbf{k}}+\text{h.c.}\right),
\end{equation}
which yields the 3D Bloch Hamiltonian $H_{\mathbf{k}}=\varepsilon_{\mathbf{k}}-2t_{\perp}\cos k_{z}-\mu$
in the bulk with a Fermi surface defined by $\varepsilon_{\mathbf{k}}-2t_{\perp}\cos k_{z}=\mu$.
Solving Eq. (\ref{eq:G-eq}) in the limit $N\to\infty$ with $p=1$,
$H_{p}^{0}=\varepsilon_{\mathbf{k}}-\mu$ and $H_{p}^{\pm}=-t_{\perp}$,
and selecting the imaginary part at complex energy $E+i\hbar/2\tau_{\mathbf{k}}$
gives the surface spectral function, 
\begin{equation}
A_{\mathbf{k}}^{\text{triv}}(E)=\frac{\text{sgn}(E+\mu-\varepsilon_{\mathbf{k}})\sqrt{R\left[4t_{\perp}^{2}-(E+\mu-\varepsilon_{\mathbf{k}})^{2}\right]}}{t_{\perp}^{2}}
\end{equation}
In writing $A_{\mathbf{k}}^{\text{triv}}(E)$, we have taken the limit
$\hbar/2\tau_{\mathbf{k}}\to0$ because the surface Green's function
lacks poles. Its only non-analyticities are square root branch cuts
when $-2|t_{\perp}|<E+\mu-\varepsilon_{\mathbf{k}}<2|t_{\perp}|$,
i.e., when $E$ lies within the bulk bands. According to (\ref{eq:sigma_general}),
the leading term in the corresponding surface conductivity is
\begin{align}
\sigma_{\text{surf}}^{\text{triv}} & =\frac{e^{2}}{2\hbar}\intop\frac{dE}{2\pi}\left(-\frac{\partial f}{\partial E}\right)\intop_{\mathbf{k}}\left(\frac{\partial\varepsilon_{\mathbf{k}}}{\partial k_{x}}\right)^{2}\frac{R\left[4t_{\perp}^{2}-(E+\mu-\varepsilon_{\mathbf{k}})^{2}\right]}{t_{\perp}^{4}}\\
 & =\frac{e^{2}\hbar}{4\pi t_{\perp}^{4}}\intop_{\mathbf{k}}v_{\mathbf{k},x}^{2}\left(\left[4t_{\perp}^{2}-(\mu-\varepsilon_{\mathbf{k}})^{2}\right]-\frac{T^{2}\pi^{2}}{6}\right)\Theta\left[4t_{\perp}^{2}-(\mu-\varepsilon_{\mathbf{k}})^{2}\right]
\end{align}
where $v_{\mathbf{k},x}=\frac{1}{\hbar}\partial_{k_{x}}\varepsilon_{\mathbf{k}}$
and we have performed a Sommerfeld expansion for $f(E)$. If there
are multiple trivial Fermi surfaces, each will contribute additively
to $\sigma_{\text{surf}}^{\text{triv}}$. The temperature dependent
part, $\Delta\sigma_{\text{surf}}^{\text{triv}}(T)\equiv\sigma_{\text{surf}}^{\text{triv}}(T)-\sigma_{\text{surf}}^{\text{triv}}(0)$
is given by
\begin{equation}
\Delta\sigma_{\text{surf}}^{\text{triv}}(T)=-\frac{e^{2}\hbar}{96\pi}\frac{T^{2}\left\langle v_{x}^{2}\right\rangle _{FS}A_{FS}}{t_{\perp}^{4}}
\end{equation}
where $A_{FS}$ is the area of the surface projection of the Fermi
surface and $\left\langle v_{x}^{2}\right\rangle =\frac{1}{A_{FS}}\int d^{2}kv_{\mathbf{k},x}^{2}$
is the mean square velocity in this area.

Thus, a trivial Fermi surface also produces a $\tau$-independent
surface conductivity whose leading $T$ dependence at low temperatures
is $\sim T^{2}$. However, the surface conductivity due to a trivial
Fermi surface is proportional to its cross-sectional area, so it vanishes
if the Fermi pocket is emptied. Physically, the dependence on $A_{FS}$
indicates that $\Delta\sigma_{\text{surf}}^{\text{triv}}(T)$ is a
surface manifestation of quasiparticle excitations above the bulk
Fermi surface. The $\mu$-dependence of $\sigma_{\text{surf}}^{\text{triv}}(T)$
is rather non-universal and depends on band structure details. In
contrast, (i) the $T$-dependent part of $\sigma_{\text{anom}}$ in
WSMs, namely, $\Delta\sigma_{\text{anom}}(T)=\frac{e^{2}}{6\hbar}\sum_{j}\frac{T^{2}v_{j,x}^{2}}{t_{\perp}^{2}|\mathbf{u}_{j}\times\mathbf{v}_{j}|}$,
is independent of doping around the Weyl node and persists even as
the Fermi surface around a Weyl node shrinks to a point, and (ii)
$\sigma_{\text{anom}}$ only depends on a particular combination of
$T$ and $\mu$, namely, $\mu^{2}+T^{2}\pi^{2}/3$. 
Both these features of $\sigma_{\text{anom}}$ are consequences
of the linear dispersion that is innate to a Weyl node.

\end{widetext}

\end{document}